\documentclass[a4paper,fleqn,usenatbib]{mnras}

\usepackage{newtxtext,newtxmath}
\usepackage[T1]{fontenc}
\usepackage{ae,aecompl}

\usepackage{graphicx}	% Including figure files
\usepackage{amsmath}	% Advanced maths commands
\usepackage{amssymb}	% Extra maths symbols

\newcommand{\angstrom}{\text{\normalfont\AA}}

\title[Sounding stellar cycles with $Kepler$ -- III.]{Sounding stellar cycles with $Kepler$ -- III. Comparative analysis of chromospheric, photometric and asteroseismic variability\footnote{Based on observations made with the Nordic Optical Telescope, 
operated on the island of La Palma jointly by Denmark, Finland, Iceland, 
Norway, and Sweden, in the Spanish Observatorio del Roque de los 
Muchachos of the Instituto de Astrof{\'i}sica de Canarias.}}

\author[C. Karoff et al.]{C.~Karoff$^{1,2}$\thanks{E-mail: karoff@phys.au.dk (CK)},
T.~S.~Metcalfe,$^{3,4}$,
B.~T.~Montet$^{5}$,
N.~E.~Jannsen$^{6,2}$,
\newauthor{A.~R.~G.~Santos$^{3}$,M.~B.~Nielsen$^{7,8}$ and W.~J.~Chaplin$^{8,2}$}  
\\
$^{1}$Department of Geoscience, Aarhus University, H{\o}egh-Guldbergs Gade 2, 8000, Aarhus C, Denmark\\
$^{2}$Stellar Astrophysics Centre, Department of Physics and Astronomy, Aarhus University, Ny Munkegade 120, 8000, Aarhus C, Denmark\\
$^{3}$Space Science Institute, 4750 Walnut Street, Suite 205, Boulder CO 80301, USA\\
$^{4}$Max-Planck-Institut f\"ur Sonnensystemforschung, Justus-von-Liebig-Weg 3, 37077, G\"ottingen, Germany\\
$^{5}$Department of Astronomy and Astrophysics, University of Chicago, 5640 S. Ellis Ave, Chicago, IL 60637, USA\\
$^{6}$Nordic Optical Telescope, Apartado 474, E-38700 Santa Cruz de La Palma, Spain\\
$^{7}$Center for Space Science, NYUAD Institute, New York University Abu Dhabi, PO Box 129188, Abu Dhabi, UAE\\
$^{8}$School of Physics \& Astronomy, University of Birmingham, Edgbaston, Birmingham B15 2TT, UK
}

\date{Accepted XXX. Received YYY; in original form ZZZ}

\pubyear{2018}

\begin{document}
\label{firstpage}
\pagerange{\pageref{firstpage}--\pageref{lastpage}}
\maketitle

\begin{abstract}
By combining ground-based spectrographic observations of variability in the chromospheric emission from Sun-like stars with the variability seen in their eigenmode frequencies, it is possible to relate the changes observed at the surfaces of these stars to the changes taking place in the interior. By further comparing this variability to changes in the relative flux from the stars, one can obtain an expression for how these activity indicators relate to the energy output from the stars. Such studies become very pertinent when the variability can be related to stellar cycles as they can then be used to improve our understanding of the solar cycle and its effect on the energy output from the Sun.

Here we present observations of chromospheric emission in 20 Sun-like stars obtained over the course of the nominal 4-year $Kepler$ mission. Even though 4 years is too short to detect stellar equivalents of the 11-year solar cycle, observations from the $Kepler$ mission can still be used to analyse the variability of the different activity indicators thereby obtaining information of the physical mechanism generating the variability. The analysis reveals no strong correlation between the different activity indicators, except in very few cases. We suggest that this is due to the sparse sampling of our ground-based observations on the one hand and that we are likely not tracing cyclic variability on the other hand. We also discuss how to improve the situation.
\end{abstract}

 
\section{Introduction}
Our Sun shows variability on a number of time scales, where the most pronounced are: the 5-minute p-mode oscillations, the 25-30 days rotational modulation and the 11-year solar cycle \citep{2017NatAs...1..612S}. While we have witnessed significant progress in our understanding of the first two phenomena over the last decades, our knowledge of the 11-year solar cycle is still deficient. Especially, our understanding of the long-term modulation of the amplitude of the 11-year solar cycle, where the occurrence of so-called grand minima, like the 17th century Maunder Minimum, constitute a huge challenge \citep{1976Sci...192.1189E}. This challenge can be met by making better observations and better models of the Sun and by comparing these to observations and models of other stars \citep{2008ssma.book.....S}. 

Most of the important information we have of the 11-year solar cycle comes from either direct observations of sunspots or from observations of standing oscillations inside the Sun through helioseismology \citep{2002RvMP...74.1073C, 2004SoPh..224..217G}. With helioseismology we have learned about the differential rotation pattern of the Sun's deep interior and about the meridional circulation in the Sun's convection zone.  

For other Sun-like stars (stars with masses and radii similar to the Sun), we have so far mainly obtained information of their magnetic cycles through observations of the temporal evolution of the emission in the Ca~{\sc ii} $H$ and $K$ lines as has been done from the Mount Wilson and Lowell observatories \citep{1995ApJ...438..269B, 2005PASP..117..657W}. Expectations where therefore high, when the $Kepler$ mission was launched on 7 March 2009. Though the nominal mission lifetime was only 3.5 years, there was in principle nothing that would prevent the mission to be operational for 10+ years, or so we thought. A 10+ years mission would have allowed us not only to detect stellar cycles in a large number of Sun-like stars, it would also have allowed us to use asteroseismology to sound the effect these cycles would have on the deep interior or vice versa. This information could be used to test and improve the models we have for the 11-year solar cycle.

In 2009 we therefore started the {\it Sounding stellar cycles with Kepler} program, where we made dual observations of 20 carefully selected Sun-like stars from both the Nordic Optical Telescope (NOT) and $Kepler$. We have earlier presented the strategy behind this program \citep[][Paper I]{2009MNRAS.399..914K} and the first results from the measurements of chromospheric emission \citep[][Paper II]{2013MNRAS.433.3227K}. Here we present the full set of measurements made over the course of the $Kepler$ mission from early 2010 to late 2014. 

As $Kepler$ was only able to point to the Cygnus field for 4 years, studies of rotation of Sun-like stars experienced more progress than studies of stellar cycles of Sun-like stars. From early on in the mission, surface rotation was identified in a large number of, mainly active, stars \citep{2013ApJ...775L..11M, 2013A&A...557L..10N, 2014ApJS..211...24M, 2014A&A...572A..34G, 2016MNRAS.461..497B, 2017A&A...605A.111C} and there were even indications of differential rotation \citep{2013A&A...560A...4R, 2014A&A...569A.113B, 2015A&A...583A..65R, 2016MNRAS.463.1740B, 2018ApJ...852...46K, 2018Sci...361.1231B, 2018arXiv181008630B}, though the possibility to accurately extract differential rotation from photometry has been called into question \citep{2015MNRAS.450.3211A, 2018ApJ...865..142B}. The main result of these measurements was that the observations agreed nicely with the theory, especially for the behaviour of differential rotation \citep{2013A&A...560A...4R}. The new observations did however, also indicate that stars with a relative main sequence age older than the Sun lose angular momentum slower than generally believed by stellar evolution theory \citep{2016Natur.529..181V}. Though this result has received some criticism, mainly focusing on selection bias and underestimation of the age uncertainties \citep{2016AN....337..810B}, the result has later been confirmed by another independent study \citep{2016A&A...592A.156D}. Lately, a number of theoretical explanations  for the small angular momentum loss of old solar-type stars have also emerged \citep{2016ApJ...826L...2M, 2017ApJ...845...79B, 2017SoPh..292..126M, 2017ApJ...848...43J}.

Based on asteroseismic analysis of observations from the CoRoT satellite, there were early claims of stellar cycles as short as 120 days \citep{2010Sci...329.1032G}. We now know that $F$-type stars slightly hotter than the Sun, with very thin outer convective zones, tend to show irregular variability with characteristic times scales of a few hundred days and not bona fide cycles \citep{2013A&A...550A..32M, 2014A&A...562A.124M, 2015A&A...583A.134F, 2016A&A...589A.103R}. Great care should thus be taken when claiming the detection of any cycles with only 4 years of $Kepler$ observations. It is however, very interesting to compare the variability in such different activity indicators, as it can teach us something about the physical mechanisms responsible for generating the variability \citep{2016A&A...596A..31S, 2017ApJ...834..207M}. Especially, it is interesting to compare magnetic activity indicators originating from the surface to indicators originating from the interior. It is therefore very fortunate that a number of recent studies have been able to use asteroseismology to trace magnetic variability \citep{2017A&A...598A..77K, 2018A&A...611A..84S, 2018ApJS..237...17S}.

As $Kepler$ was designed for exoplanet detection, the photometry is not suited for identifying stellar activity cycles or the effect of stellar activity. The main reason for this is that the mean level of the photometry is normalised by the calibration method every three months. A solution to this was found by \citet{2017ApJ...851..116M}, who employed the so-called full-frame images for reconstruction of the photometric long-term variability. They used these measurements to show a transition from spot-dominated to faculae-dominated variability for rotation periods between 15 and 25 days. Another solution was found by \citet{2017A&A...603A..52R}, who used the variability rather than the intensity to search for stellar cycles in active solar-type stars. Both studies suffered from the short lifetime of the $Kepler$ mission and were thus not able to identify bona fide stellar cycle variability in any of their targets.

The issue with the short lifetime of the $Kepler$ mission was however, not a problem for the high metallicity Sun-like star KIC 8006161. This star was not only part of our NOT program, it was also observed as part of the Mount Wilson HK project \citep{1995ApJ...438..269B} and the California Planet Search program \citep{2005PASP..117..657W}. We were thus able to reconstruct a time series of the chromospheric activity measurements dating back to the mid-nineties (with even a few data points in the late seventies and early eighties) that allowed us to measure a bona fide cycle period of $7.41\pm1.16$ year. As the rising phase of the last cycle was covered by observations from the $Kepler$ mission, this allowed us to compare the realisation of the magnetic activity in different activity indicators related to the chromosphere, photosphere and interior. The relation between these indicators agreed very nicely with theoretical predictions for this high metallicity star, indicating that the cycle in KIC 8006161 does indeed have the same nature as the solar cycle \citep{2018ApJ...852...46K}.

This study continues the work in Paper I and Paper II. In Paper I we presented the idea behind the {\it Sounding stellar cycles with Kepler} program and discussed the strategy for selecting targets. In Paper II we presented the first measurements of chromospheric emission and analysed the dependency between the mean values of these measurements and fundamental stellar parameters obtained from asteroseismology. At the time of the submission and publication, we expected that $Kepler$ would continue the observations in the Cygnus field for 10+ years. Shortly after the publication of Paper II the second reaction wheel failed and the Cygnus field was abandoned. As the idea from the beginning of the program was that we would only make simultaneous observations with the NOT as long as $Kepler$ was operational, we thus also interrupted our NOT program -- meaning that we stopped submitting proposals. In this study we thus present 4 years of simultaneous observations of magnetic activity related variability in our 20 target stars. 

The main scope of this study is to present the measurements of chromospheric variability from the NOT made simultaneously with the observations by $Kepler$. In order to evaluate the information content in the measurements we also compare our measurements of chromospheric activity variability with asteroseismic measurements \citep{2018ApJS..237...17S} and photometric measurements \citep{2017ApJ...851..116M}. 

Though we did interrupt our NOT program when the second reaction wheel on $Kepler$ failed, we have continued working on a solution for multidecadal observations of the 20 targets, with a dedicated spectrograph installed at the Hertzsprung Stellar Observations Network Group (SONG) telescope at Teide Observatory \citep{2017ApJ...836..142G}. We will provide a brief overview of this idea and discuss the lesson learned from our NOT program with respect to S/N, sampling and observation time span.

\section{Observations}
The spectrographic observations are described in Paper II. They were obtained with the high-resolution FIbre-fed Echelle Spectrograph (FIES) mounted on the NOT \citep{2000mons.proc..163F}. 
The target list is provided in Table 1. Most stars were observed for 3 epochs per year in 2010, 2011, 2012 and 2013 (12 epochs in total). The low-resolution fibre (R=25,000) was used for the observations. The spectra were obtained with 7-minute exposures resulting in a S/N above 100 at the blue end of the spectrum for the faintest starts. 

The spectra were reduced as described in Paper II using FIEStool\footnote{http://www.not.iac.es/instruments/fies/fiestool/FIEStool.html}.

\begin{table*}
\caption{Target list for the {\it sounding stellar cycles with Kepler} programme. We also list the Kepler magnitude and $B-V$ values from \citet{2000A&A...355L..27H} and [Fe/H] from \citet{2012MNRAS.423..122B}. The uncertainty on [Fe/H] is 0.06 dex.}
\centering
\begin{tabular}{lcccccccccc}
\hline \hline
KIC ID & $\alpha$ (2000) & $\delta$ (2000) & $k_p$ & $B-V$  & [Fe/H] & $\sigma S$ & $\sigma I$ (ppm) & $\sigma ( \delta \nu)$ ($\mu$Hz)\\
\hline
01435467 & 19:28:19.84 & 37:03:35.3 & 8.9 & 0.47$\pm$0.02 & -0.01 & 0.011 & 634 & 0.24\\ 
02837475 & 19:10:11.62 & 38:04:55.9 & 8.4 & 0.43$\pm$0.02 & -0.02 & 0.018 & 617 & 0.34\\
03733735 & 19:09:01.92 & 38:53:59.6 & 8.4 & 0.41$\pm$0.02 & -0.04 & 0.007 & 1057 & --\\
04914923 & 19:16:34.88 & 40:02:50.1 & 9.4 & 0.62$\pm$0.03 & 0.17 & 0.015 & 658 & 0.07\\
06116048 & 19:17:46.34 & 41:24:36.6 & 8.4 & 0.57$\pm$0.01 & -0.24 & 0.013 & 1919 & 0.11\\ 
06603624 & 19:24:11.18 & 42:03:09.7 & 9.0 & 0.76$\pm$0.03 & 0.28 & 0.018 & 843 &0.04\\
06933899 & 19:06:58.34 & 42:26:08.2 & 9.6 & 0.59$\pm$0.04 & 0.02 & 0.019 & 736 & 0.059\\
07206837 & 19:35:03.72 & 42:44:16.5 & 9.7 & 0.46$\pm$0.06 & 0.14 & 0.020 & 1354 & 0.227\\
08006161 & 18:44:35.14 & 43:49:59.9 & 7.3 & 0.87$\pm$0.01 & 0.34 & 0.018 & 1941 & 0.276\\
08379927 & 19:46:41.28 & 44:20:54.7 & 6.9 & 0.58$\pm$0.01 & & 0.009 & 1316 & 0.161\\
08694723 & 19:35:50.58 & 44:52:49.8 & 8.8 & 0.48$\pm$0.02 & -0.59 & 0.014 & 1741 & 0.116\\
09098294 & 19:40:21.20 & 45:29:20.9 & 9.7 & 0.68$\pm$0.08 & -0.13 & 0.017 & 1246 & 0.071\\
09139151 & 18:56:21.26 & 45:30:53.1 & 9.1 & 0.52$\pm$0.03 & 0.11 & 0.013 & 1100 & 0.124\\
09139163 & 18:56:22.12 & 45:30:25.2 & 8.3 & 0.49$\pm$0.01 & 0.15 & 0.009 & 1198 & 0.287\\
10124866 & 18:58:03.46 & 47:11:29.9 & 7.9 & 0.57$\pm$0.02 & -0.30 & 0.015 & 756 & --\\  
10454113 & 18:56:36.62 & 47:39:23.0 & 8.6 & 0.52$\pm$0.02 & -0.06 & 0.009 & 1151 & 0.221\\
11244118 & 19:27:20.48 & 48:57:12.1 & 9.7 & 0.78$\pm$0.05 & 0.35 & 0.016 & 1312 & --\\
11253226 & 19:43:39.62 & 48:55:44.2 & 8.4 & 0.39$\pm$0.02 & -0.08 & 0.011 & 1307 & 0.225\\
12009504 & 19:17:45.80 & 50:28:48.2 & 9.3 & 0.55$\pm$0.03 & -0.09 & 0.012 & 952 & 0.138\\
12258514 & 19:26:22.06 & 50:59:14.0 & 8.0 & 0.59$\pm$0.01 & 0.04 & 0.012 & 1025 & 0.093\\
%Check number in table
%Mark correlation of most active stars
\hline
\end{tabular}
\label{tab1}
\end{table*}

\section{Analysis}
The most common activity indicator utilising the Ca~{\sc ii} $H$ and $K$ lines is the so-called $S$ index \citep{1991ApJS...76..383D}:
\begin{equation}
S=\alpha \cdot \frac{H+K}{R+V}
\end{equation}
where $H$ and $K$ are the recorded counts in 1.09 {\AA} full-width at half-maximum triangular bandpasses centred on the Ca~{\sc ii} $H$ and $K$ lines at 396.8 and 393.4 nm, respectively. $V$ and $R$ are two 20 {\AA} wide reference bandpasses centred on 390.1 and 400.1 nm, respectively, while $\alpha$ is a normalisation constant. 

The normalisation constant ($\alpha$) can be obtained by measuring a number of stars that were part of the Mount Wilson HK project  \citep{1991ApJS...76..383D}. The calibration does not have to be linear \citep{2010ApJ...725..875I}. This approach was however, not followed in Paper II as there was only one star in common between the Mount Wilson HK project and our targets. Instead the excess flux $\Delta\mathcal{F}_{\rm Ca}$ was used as the activity indicator. The excess flux is defined as the surface flux arising from magnetic sources, which is calculated as the flux in the Ca~{\sc ii} $H$ and $K$ lines, subtracting the photospheric flux and the so-called basal flux.

In \citet{2018ApJ...852...46K} we however, identified 21 stars that were observed by both the California Planet Search program \citep{2005PASP..117..657W, 2010ApJ...725..875I} and by some of our programs at the NOT (including both the targets in Paper I and Paper II). These stars were used by \citet{2018ApJ...852...46K} to calculate a linear transformation between the instrumental $S$ indices measured with the NOT and the $S$ indices measured as part of the California Planet Search program. In order to minimise numerical effects in the calculation of the S indices, all spectra from the NOT and Keck telescope were reanalysed using the same code. The uncertainties of the NOT measurements were obtained using nights with multiple observations to obtain the following relation between the uncertainty of the mean value of the chromospheric activity measured that night and S/N: $\sigma = 0.011/\sqrt{\rm S/N}$. An additional flat noise term of 0.002 was added in quadrature \citep{2011arXiv1107.5325L}.

Our calibrated $S$ indices can be used to calculate the excess flux \citep{1984A&A...130..353R}:
\begin{equation}
 \mathcal{F}_{\textup{1\angstrom}}=10^{-14}S\cdot C_{\rm cf}T_{\rm eff}^4,
\end{equation}
where $C_{\rm cf}$ is the conversion factor given in \citep{1982A&A...107...31M}:
 \begin{equation}
\mathrm{log}~C_{cf}=0.25(B-V)^3-1.33(B-V)^2+0.43(B-V)+0.24
 \end{equation}
 
A number of asteroseismic estimates are available for the effective temperatures, but we recommend the effective temperatures provided in \citet{2017A&A...601A..67C}
 
Our calibrated $S$ indices can also be used to calculate the so-called $R'_{\rm HK}$ activity indicator, defined as: 
\begin{equation}
R'_{\rm HK}=R_{\rm HK}-R_{\rm phot},
\end{equation}
where $R_{\rm phot}$ is the photospheric contribution to the indicator, which can be calculated using the $B-V$ colour index \citep{1984ApJ...279..763N}:
\begin{equation}
\mathrm{log}~R_{\rm phot} = -4.898+1.918(B-V)^2-2.893(B-V)^3
\end{equation}
and $R_{\rm HK}$ is calculated based on the $S$ indices and the conversion factor given above:
\begin{equation}
R_{\rm HK}=1.34\cdot10^{-4}C_{cf} S
\end{equation}

After various tests we decided to use the $S$ index as the activity indicator. Generally, the results are not significantly affected depending on if the $S$ index or $R'_{\rm HK}$ is used as the activity indicator. The reason for this is that our 20 targets span a rather small range in effective temperature. Also, the focus of this study is the temporal variability of individual stars, so the calibration is a minor issue. 

\section{Results}
The Sun shows a strong direct correlation between chromospheric activity, photometric flux and eigenmode frequencies \citep[see][and reference herein]{2018ApJ...852...46K}. The reason for this is that changes in all three parameters are caused by magnetic flux tubes rising up through the Sun. In the
outermost regions of the near surface convection zone they change the turbulent velocity affecting the eigen frequencies \citep{2004ApJ...600..464D, 2005ApJ...625..548D}, in  the photosphere they lead to dark sunspots and bright faculae and in the chromosphere they lead to plages \citep[see][for a recent review]{2013ARA&A..51..311S}. For KIC 8006161 we were able to measure a similar effect for the rising phase of the cycle that started in 2010. We therefore search for similar correlations for the remaining 19 targets. This is done by comparing our measurements of chromospheric emission to the relative flux of the stars measured in the FFI with the method by \citet{2017ApJ...851..116M} with a method similar to what was done for KIC 8006161 by \citet{2018ApJ...852...46K}. In this work, we include a larger region to search for suitable reference targets, encompassing 125 pixels from the observed image of the star on the detector, including its bleed trail, rather than 125 pixels from the stellar centroid. We note that for three targets, KIC 2837475, 6116048, and 10124866, faint background stars overlap with the PSF of the target star, which, if variable, may affect the photometry at the ppm level. The measurements of chromospheric emission is also compared to the eigenmode frequencies presented in \citet{2018ApJS..237...17S}. We use the frequency shifts calculated as the mean of the 5 central orders of the radial and dipolar modes \citep[see][for detailed description of this]{2018ApJS..237...17S}. We present the comparison in Fig.~1. Three stars, KIC 3733735, KIC 10124866 and KIC 11244118 were not included in the analysis by \citet{2018ApJS..237...17S}. The comparison between the chromospheric emission and the relative flux of these stars is presented in Fig.~2.

\begin{figure*}
\centering
\hbox{\includegraphics[width=5.8cm]{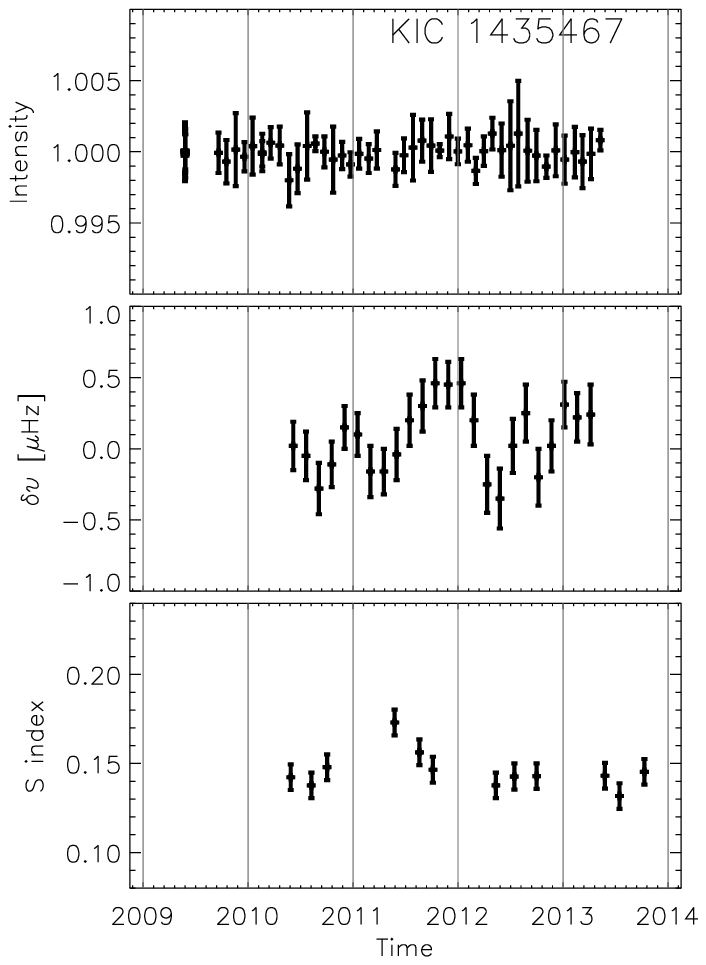}
   \hspace{0cm}\includegraphics[width=5.8cm]{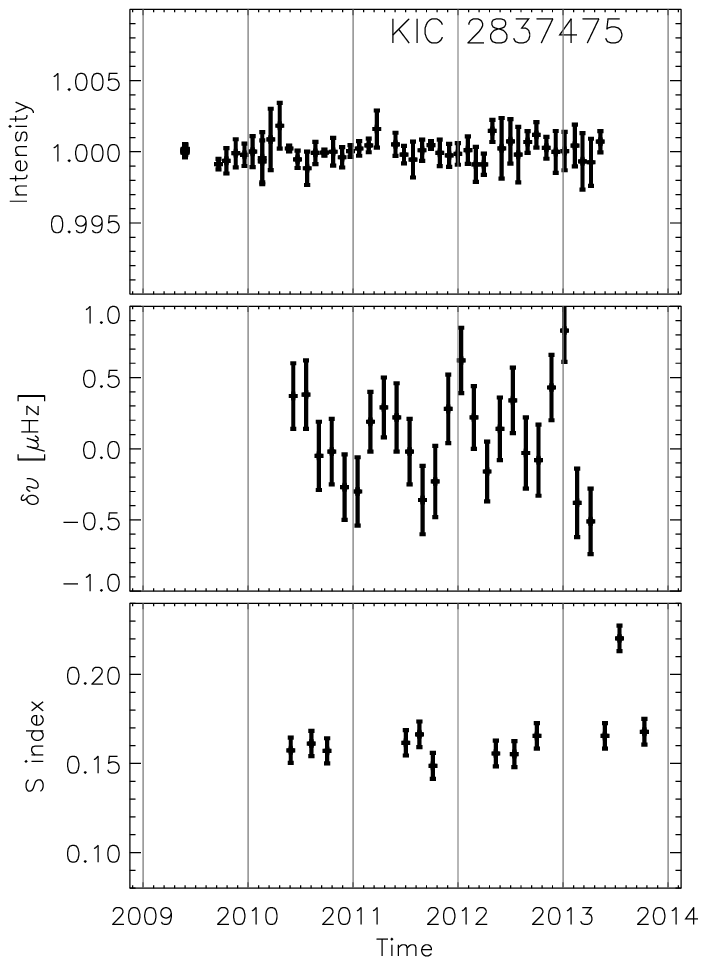}
   \hspace{0cm}\includegraphics[width=5.8cm]{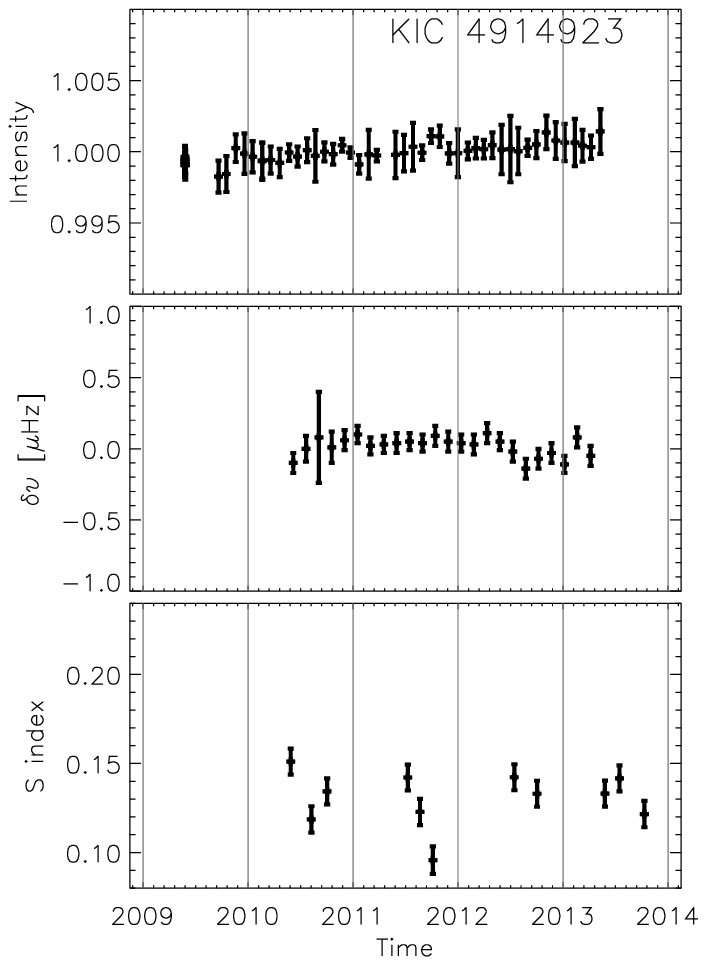}}
\vspace{0cm}
\hbox{\hspace{0cm}\includegraphics[width=5.8cm]{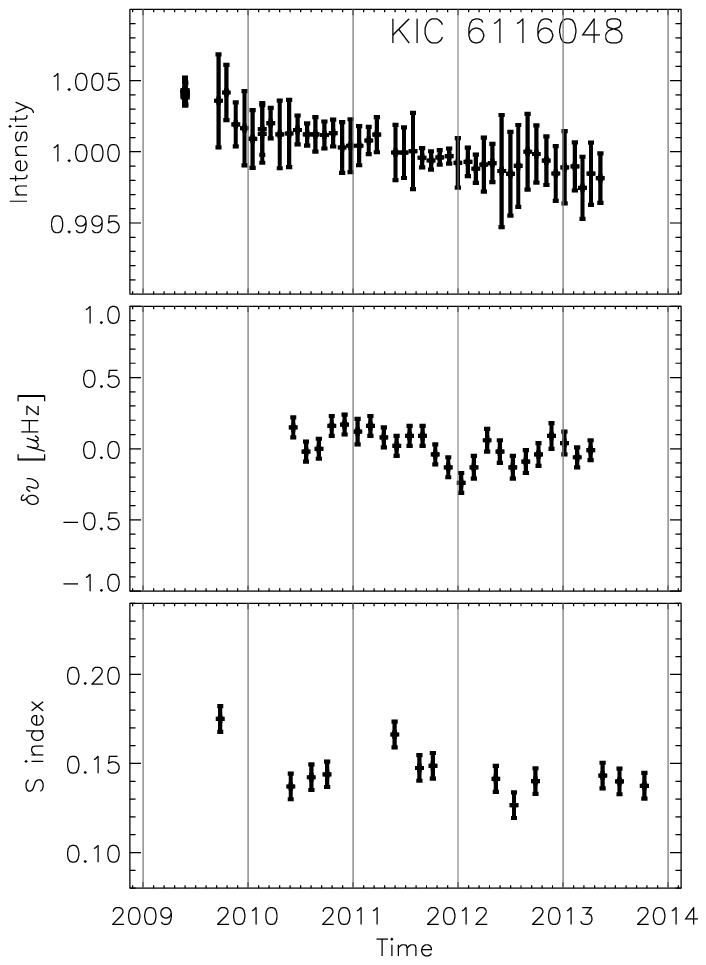}
	\includegraphics[width=5.8cm]{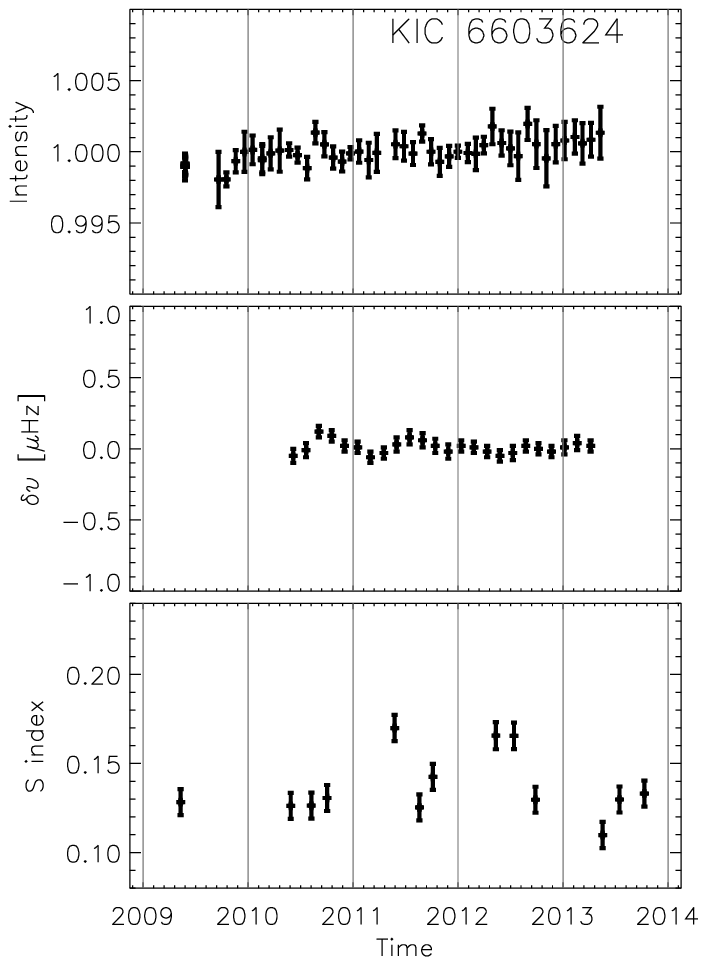}
   \hspace{0cm}\includegraphics[width=5.8cm]{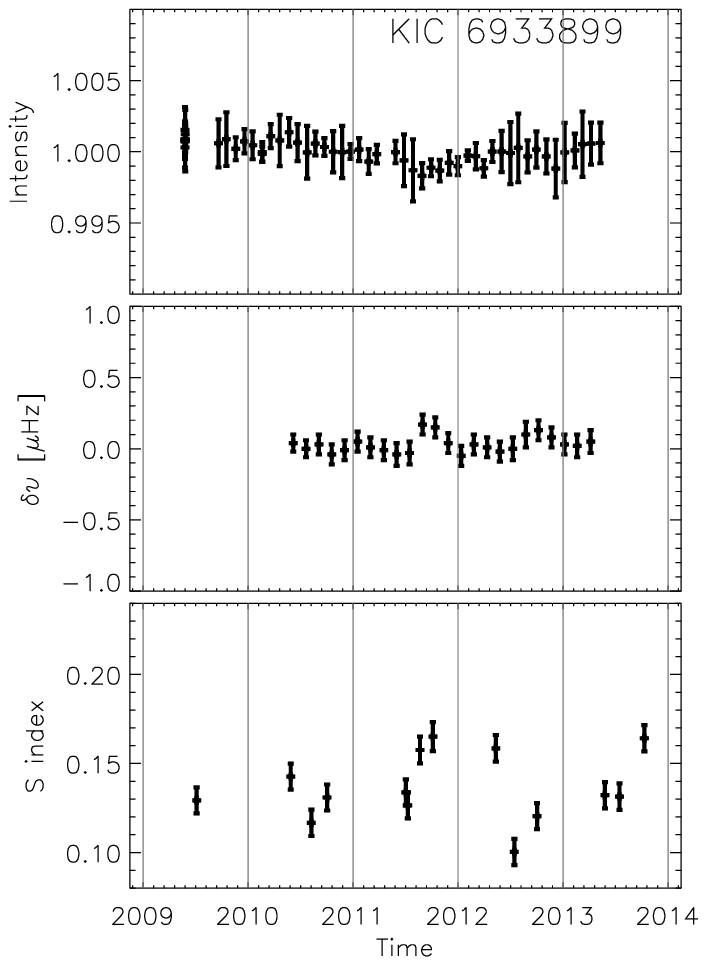}}
\vspace{0cm}
\hbox{   \hspace{0cm}\includegraphics[width=5.8cm]{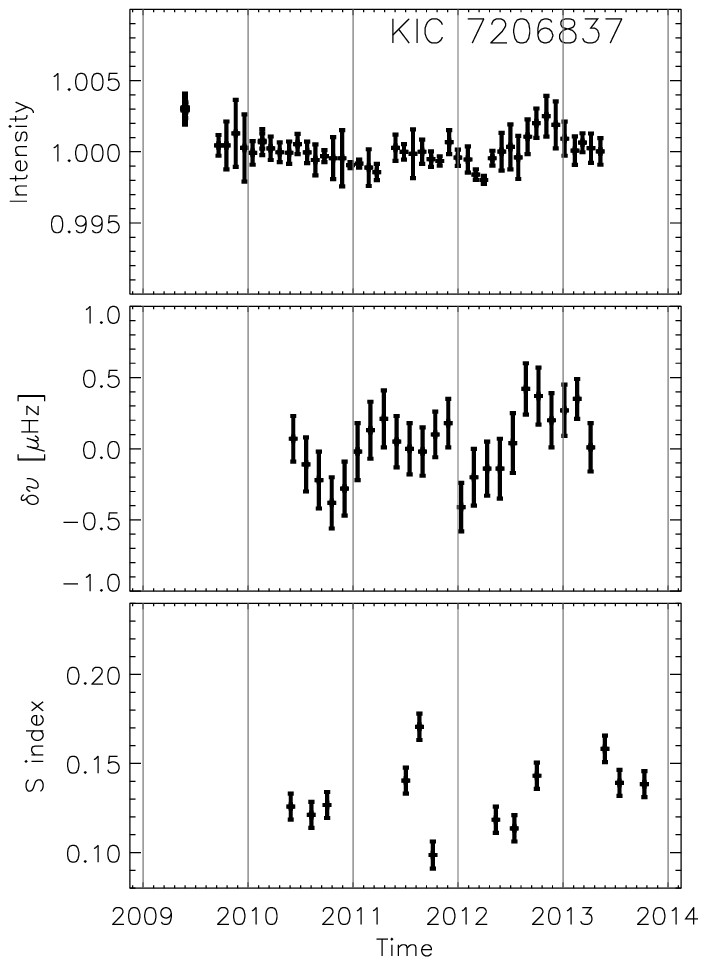}
   \hspace{0cm}\includegraphics[width=5.8cm]{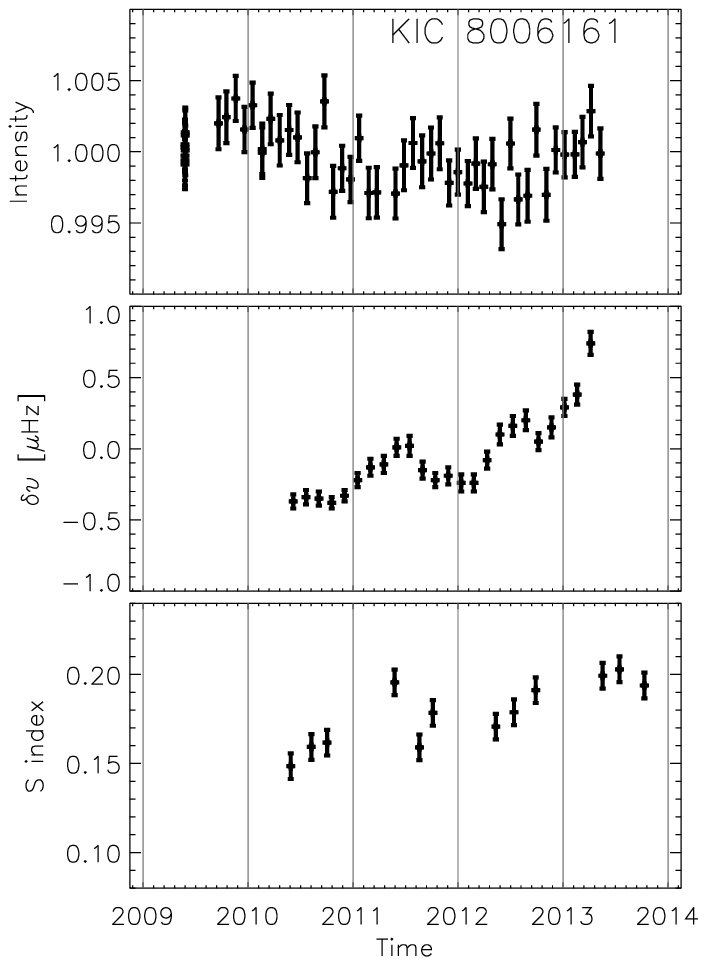}
   \includegraphics[width=5.8cm]{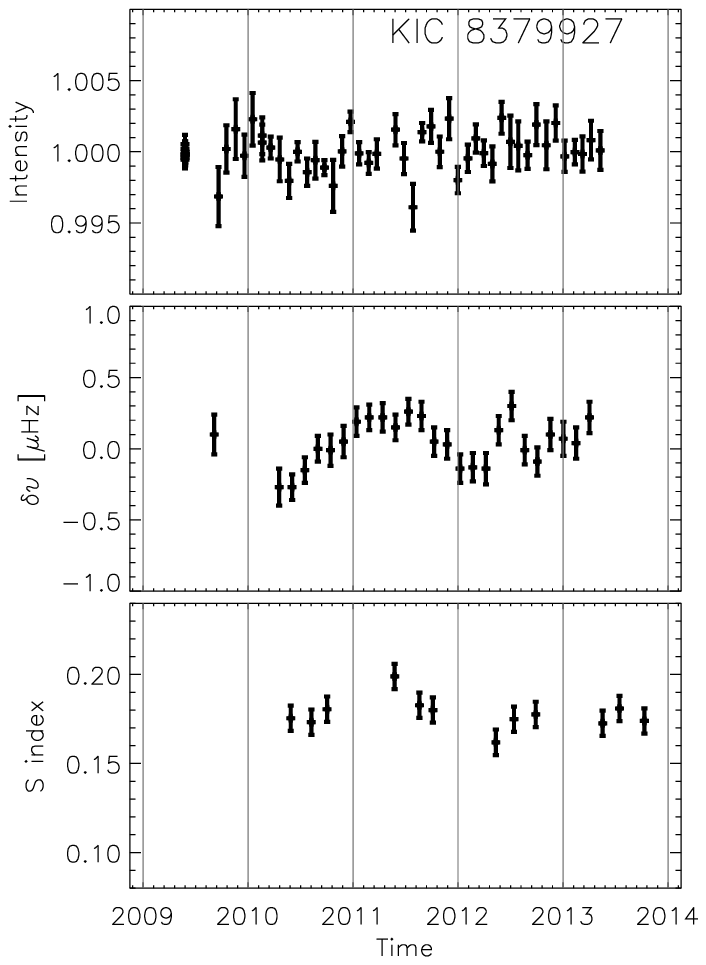}}
 \vspace{0cm}
\caption{Comparison of chromospheric activity, photometric flux and eigenmode frequencies shifts for the 17 stars with measured eigenmode frequencies in \citet{2018ApJS..237...17S}. Tables with the measured parameters are provided in the supplementary material.}
\end{figure*}
\addtocounter{figure}{-1}
\begin{figure*}
\centering
\hbox{ \hspace{0cm}\includegraphics[width=5.8cm]{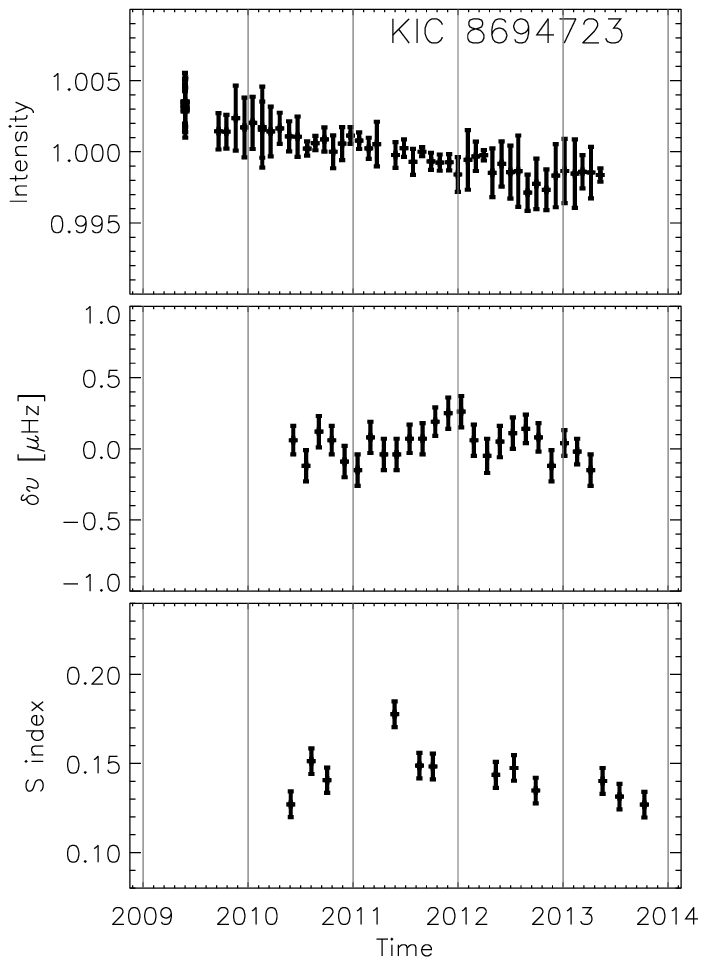}
   \hspace{0cm}\includegraphics[width=5.8cm]{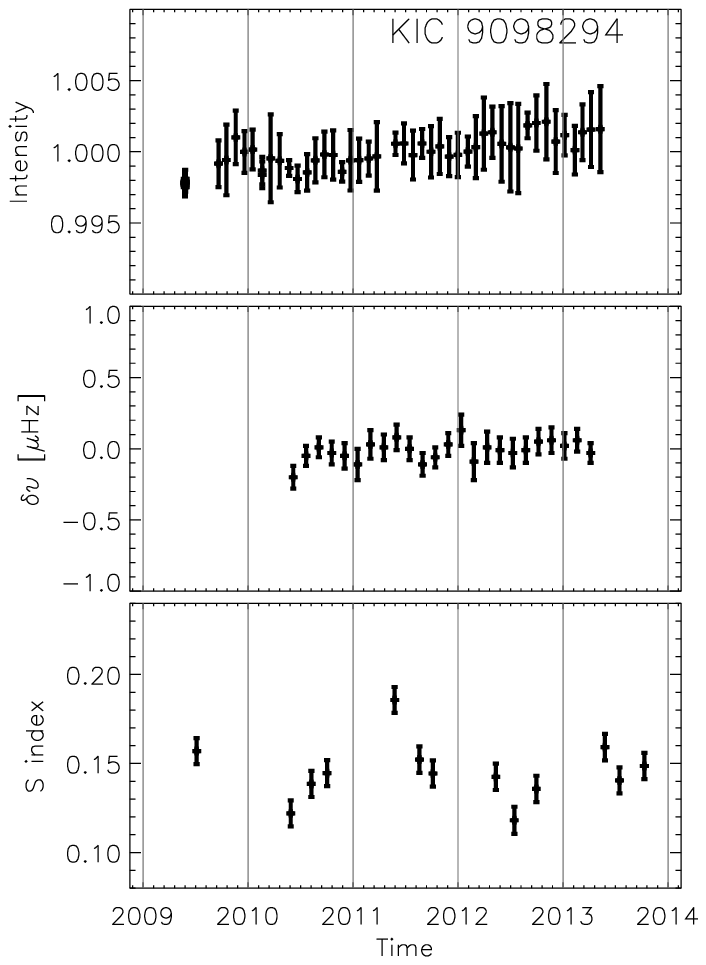}
   \hspace{0cm}\includegraphics[width=5.8cm]{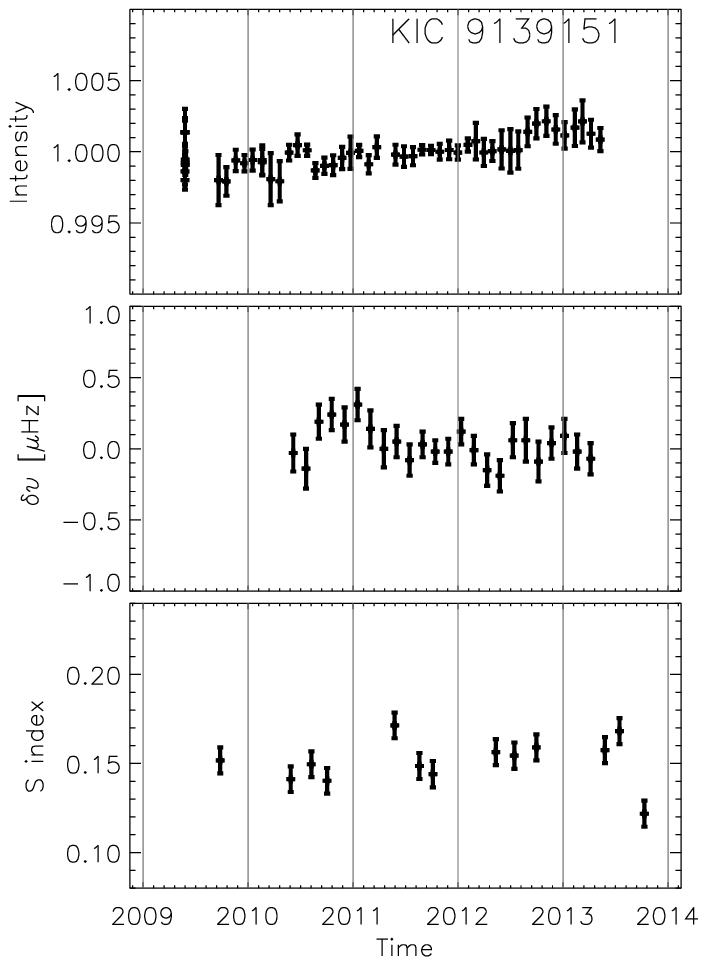}}
\vspace{0cm}
  \hbox{\includegraphics[width=5.8cm]{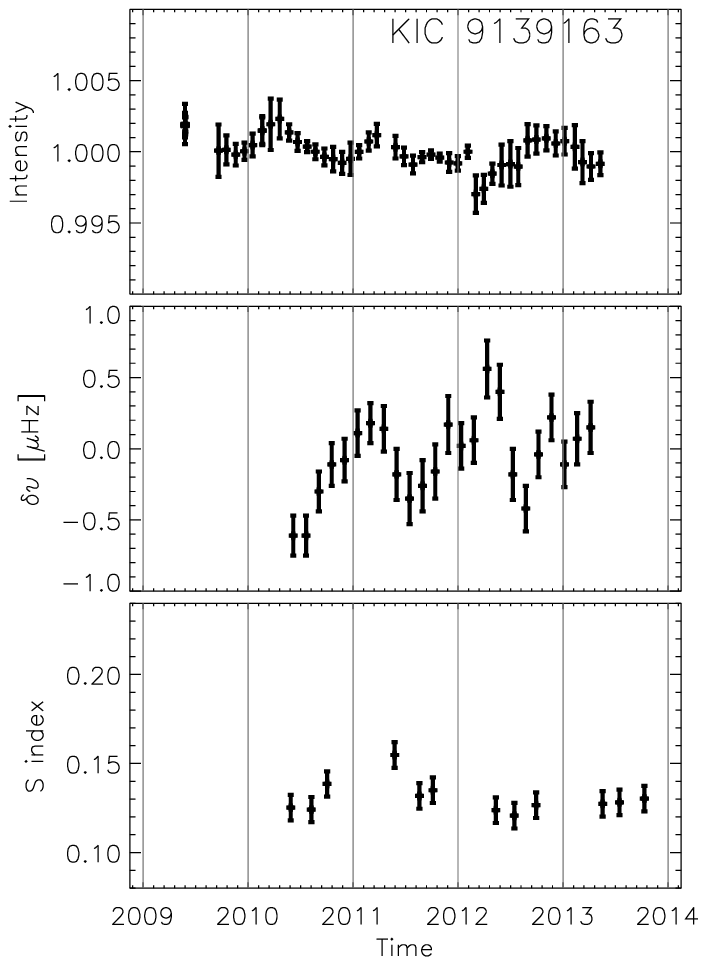}
   \hspace{0cm}\includegraphics[width=5.8cm]{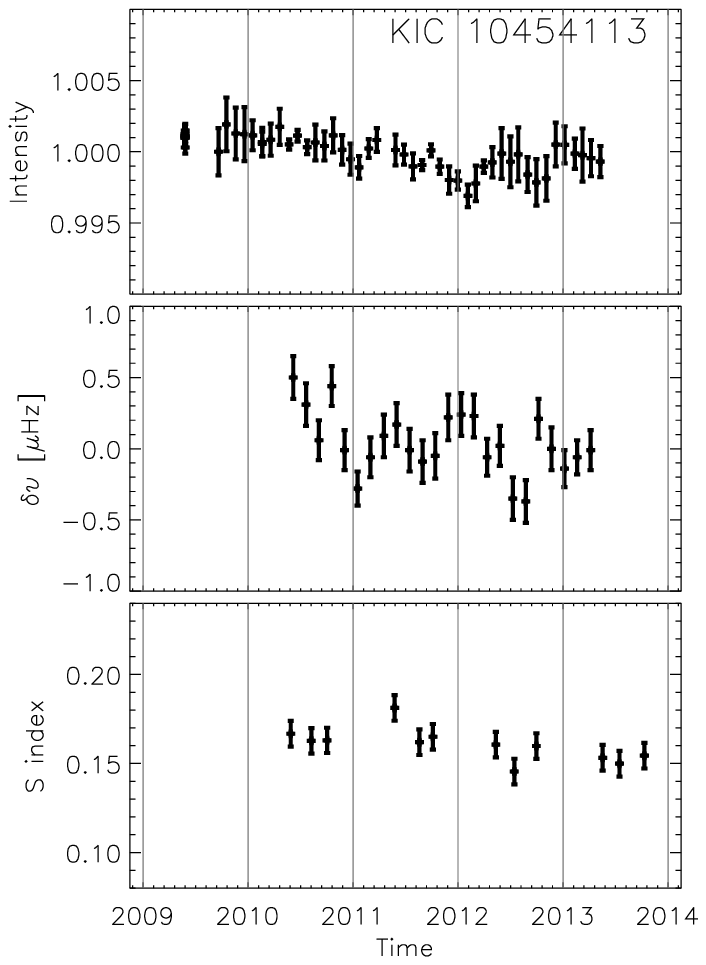}
   \hspace{0cm}\includegraphics[width=5.8cm]{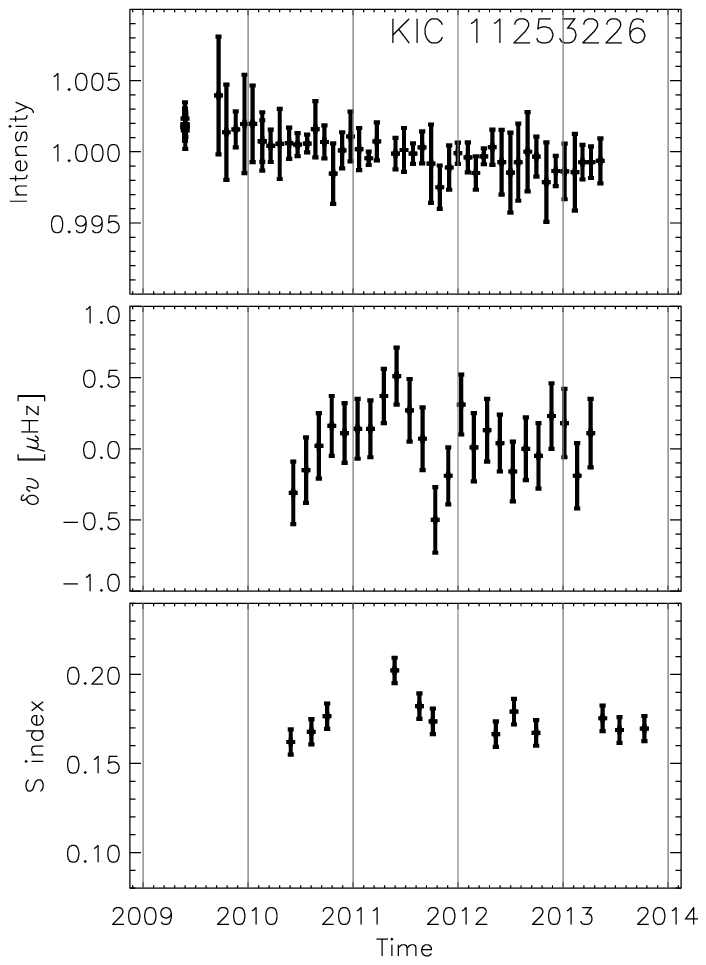}}
\vspace{0cm}
  \hbox{\hspace{0cm}\includegraphics[width=5.8cm]{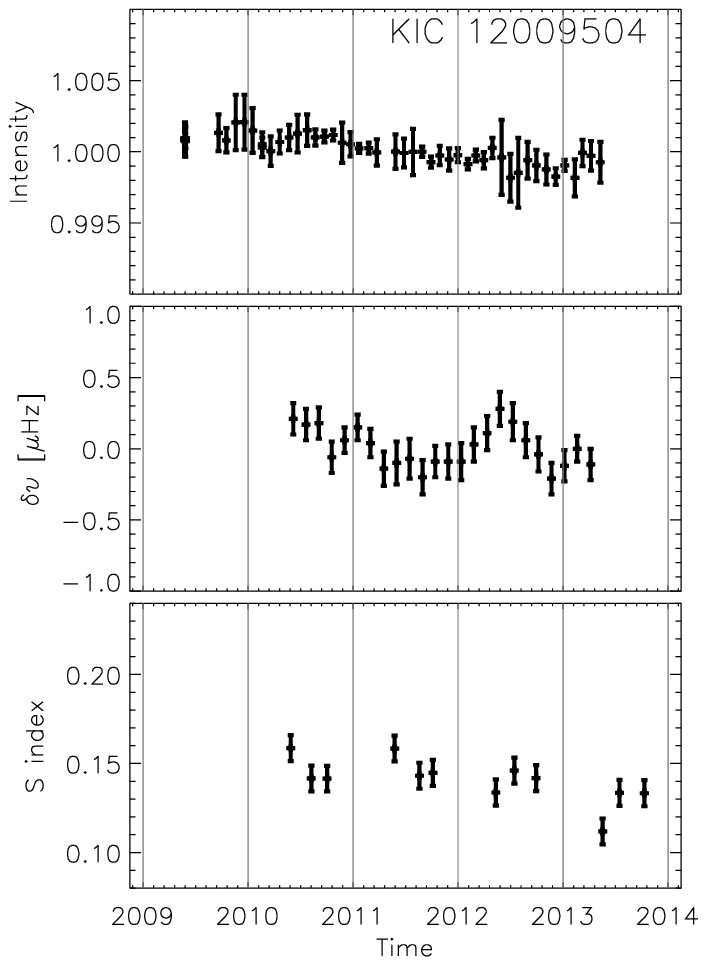}
  	\hspace{0cm}\includegraphics[width=5.8cm]{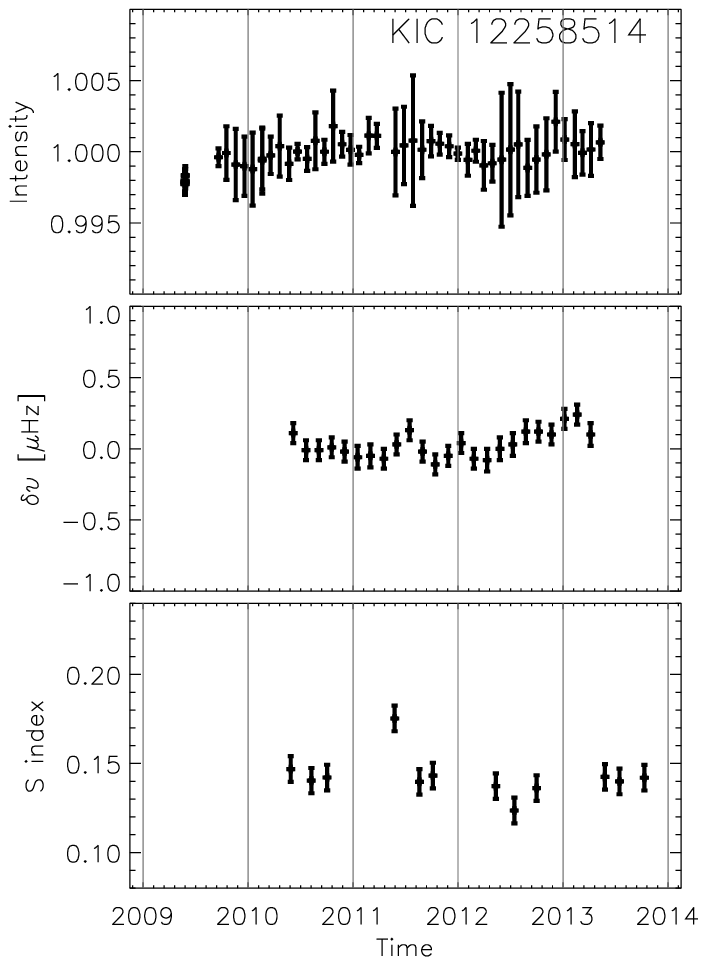}}
\caption{continued}
\end{figure*}

\begin{figure*}
\centering
\hbox{\includegraphics[width=5.8cm]{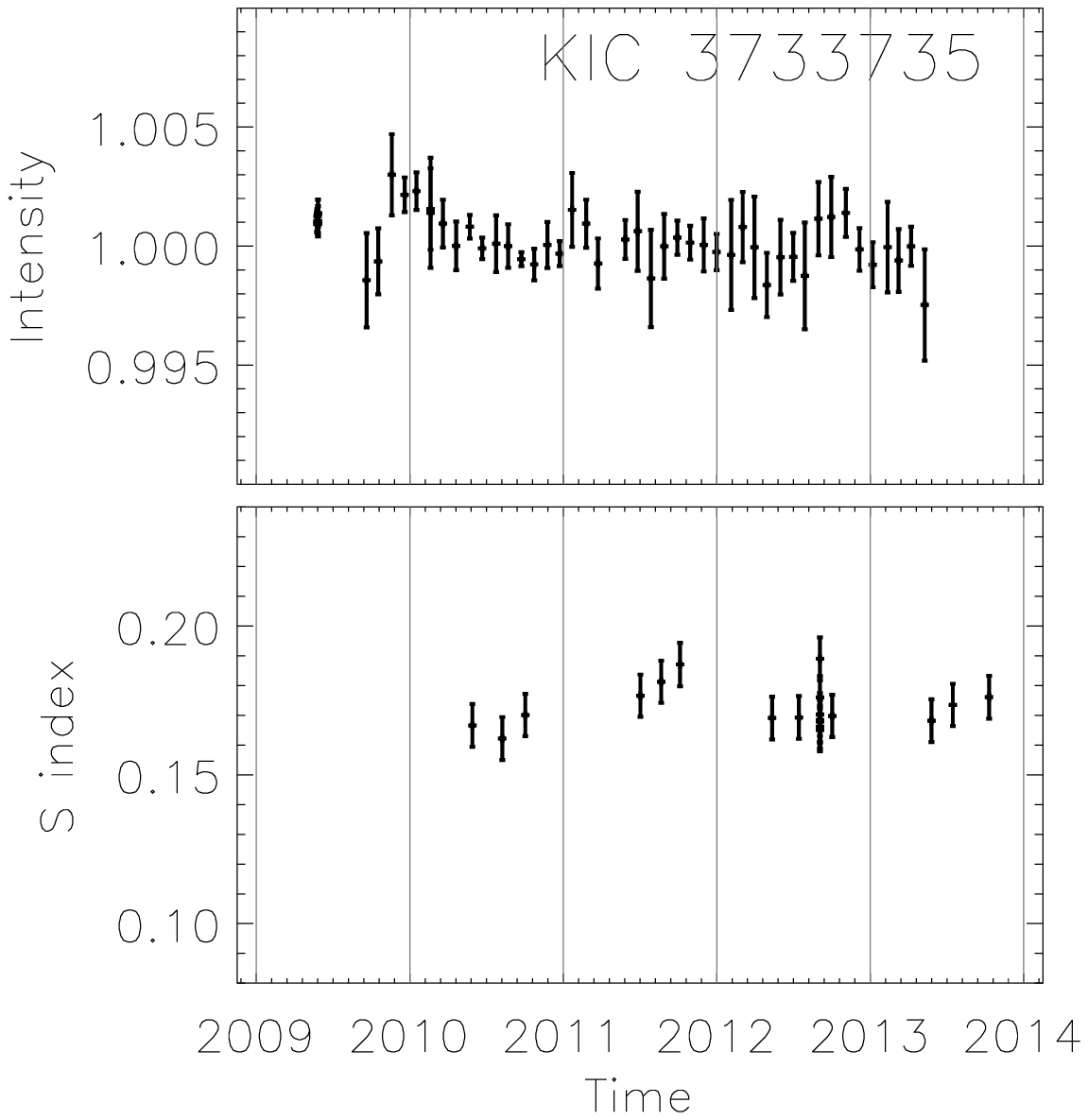}
   \hspace{0cm}\includegraphics[width=5.8cm]{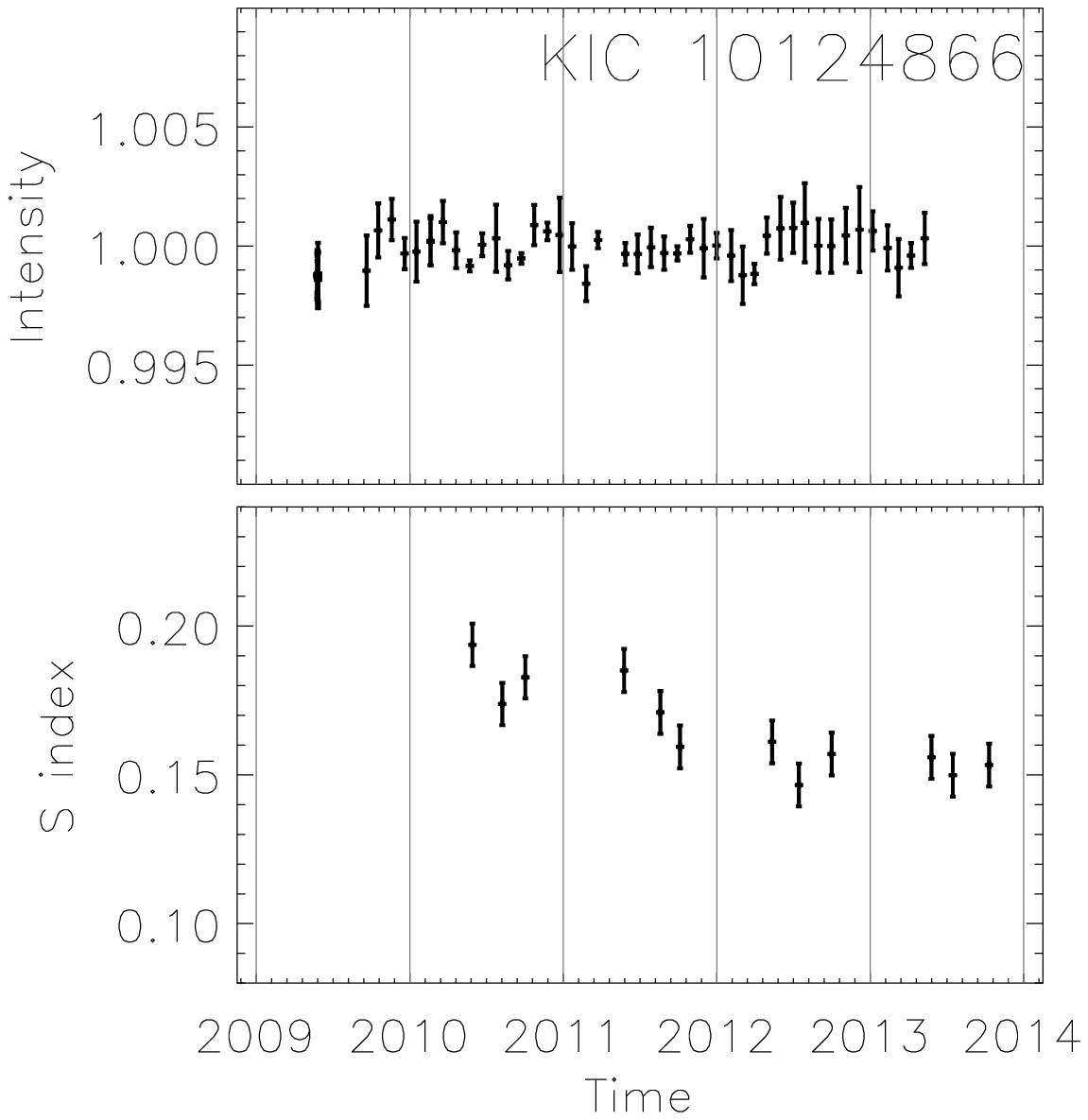}
	\vspace{0cm}\includegraphics[width=5.8cm]{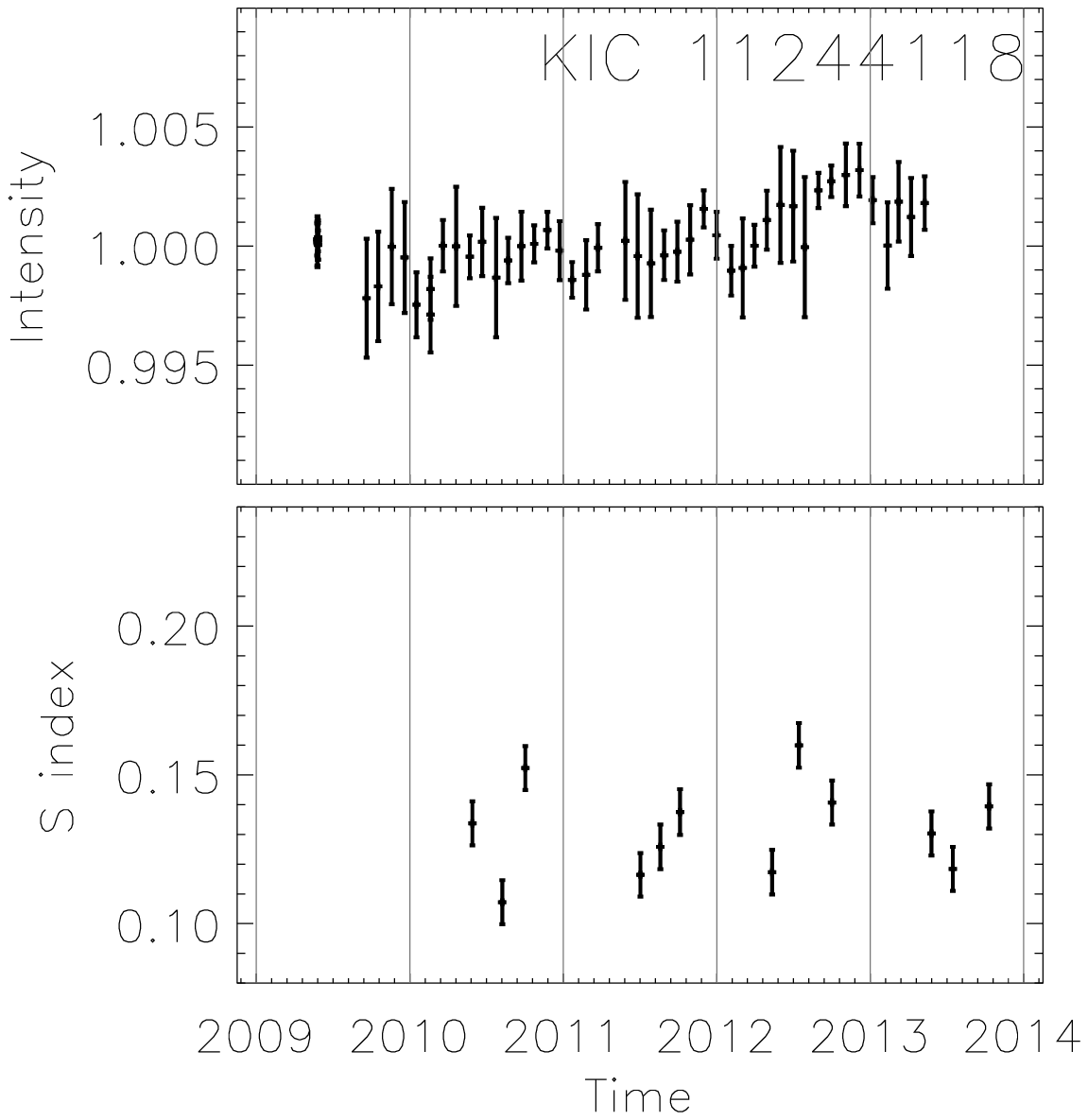}}
\caption{Comparison of chromospheric activity and photometric flux for the 3 stars without measured eigenmode frequencies in \citet{2018ApJS..237...17S}. Tables with the measured parameters are provided in the supplementary material.}
\end{figure*}

Based on the results presented in Figs.~1 \& 2 we have calculated the standard deviations of each time series. These are provided in Table~1 and in Figs.~3--5 we show the correlation between these standard deviations of the different activity indicators.

We have also calculated the correlation between the chromospheric emission and the eigenmode frequencies (Fig.~6) and between the chromospheric emission and the relative flux (Fig.~7).

\begin{figure}
\includegraphics[width=8cm]{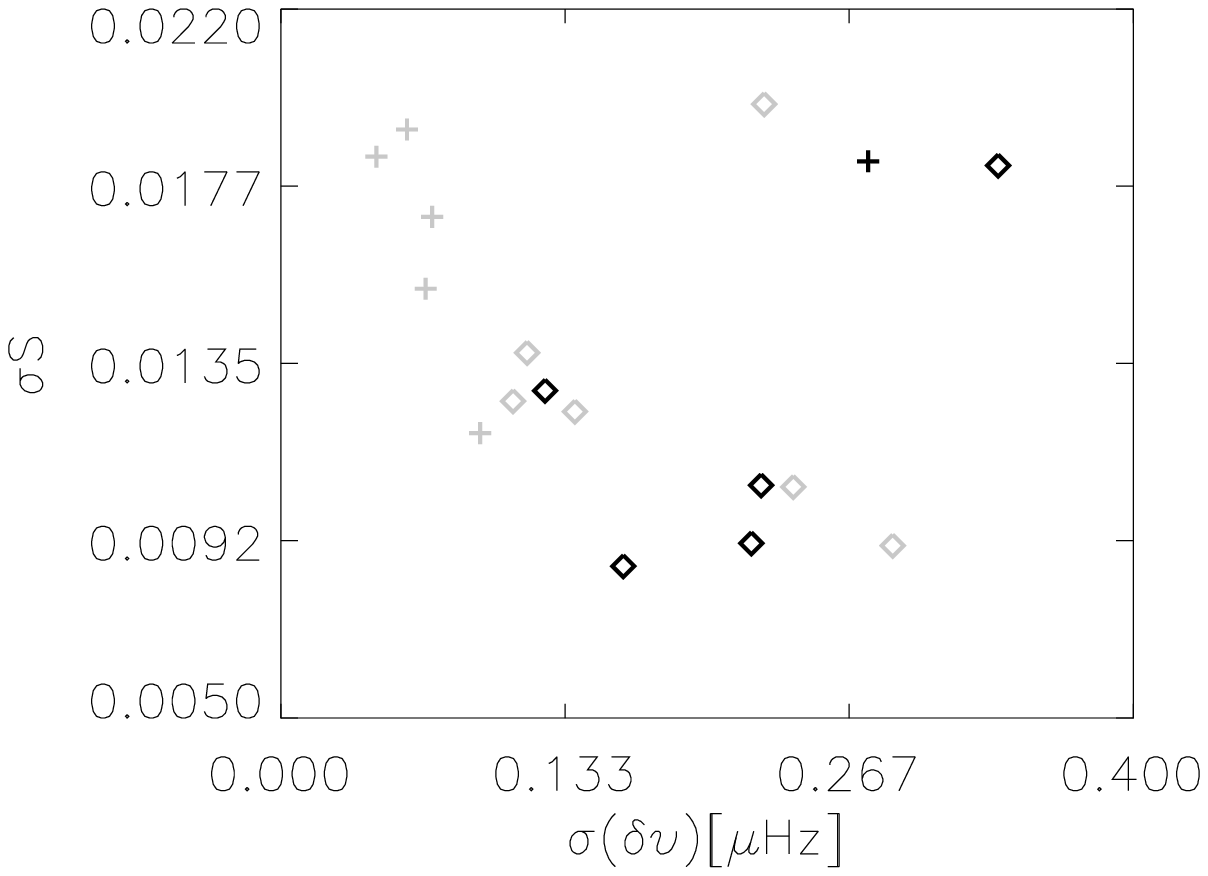}
\caption{Relation between the standard deviation of the eigenmode frequencies $\sigma(\delta \nu)$ and the chromospheric emission $\sigma S$. Gray make stars with a $S$ index less than 0.15. Crosses mark G-type dwarfs and diamonds mark F-type dwarfs. The spectral type classification has been made based on the effective temperatures.}
\end{figure}

\begin{figure}
\includegraphics[width=8cm]{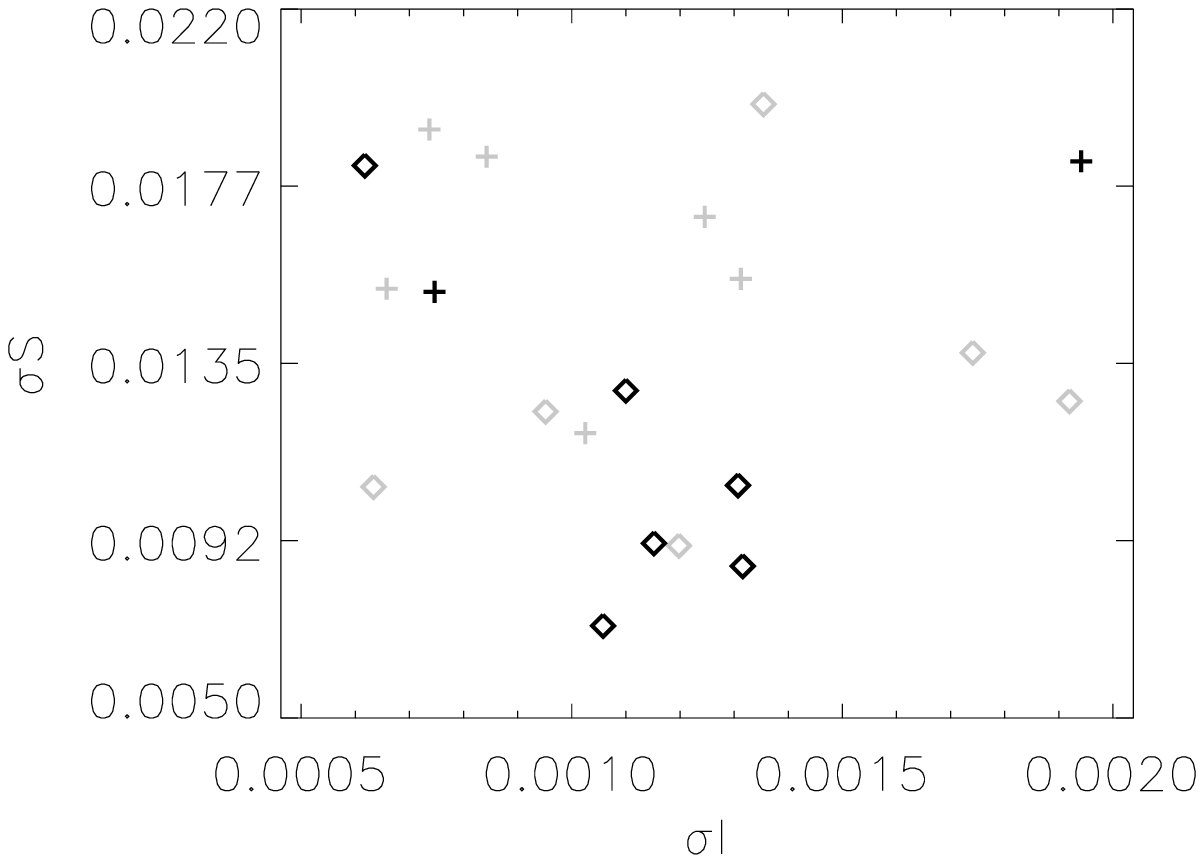}
\caption{Relation between the standard deviation of the relative flux $\sigma I$ and the chromospheric emission $\sigma S$. Symbols as in Fig.~3.}
\end{figure}

\begin{figure}
\includegraphics[width=8cm]{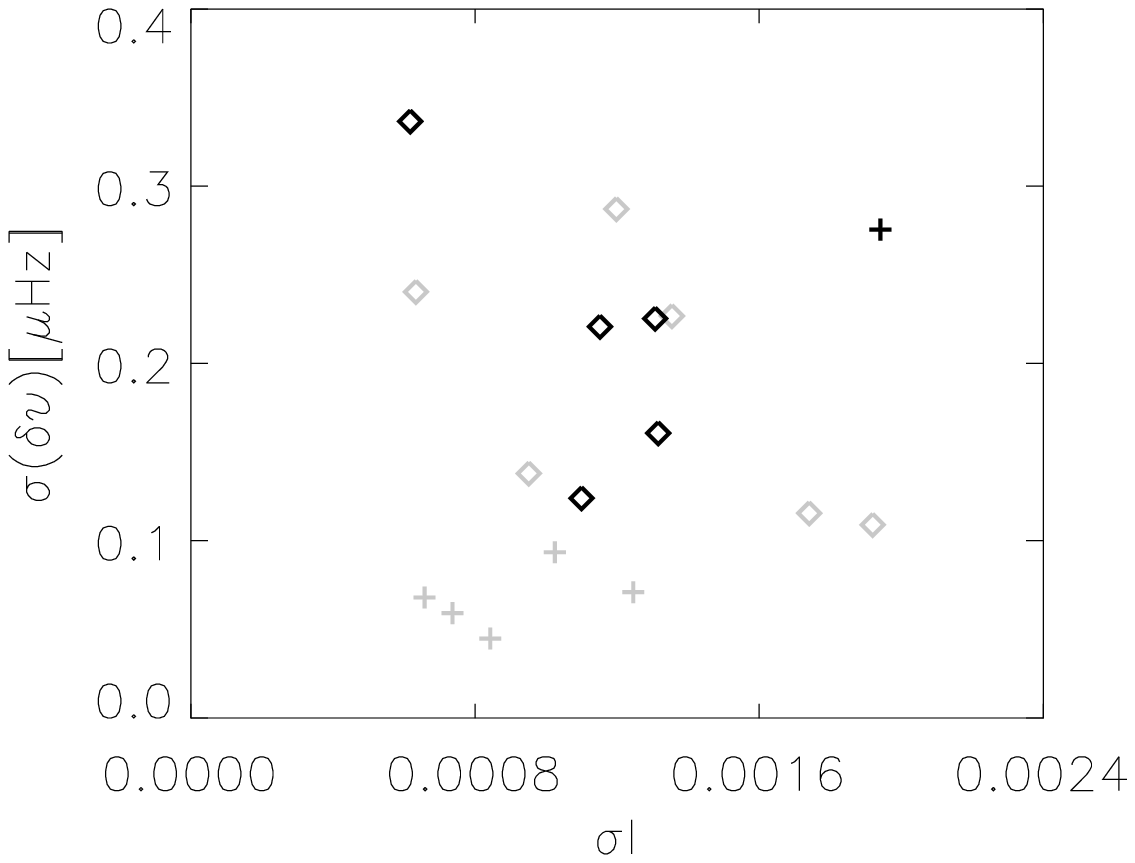}
\caption{Relation between the standard deviation of the relative flux $\sigma I$ and the eigenmode frequencies $\sigma(\delta \nu)$. Symbols as in Fig.~3.}
\end{figure}

\begin{figure}
\includegraphics[width=8cm]{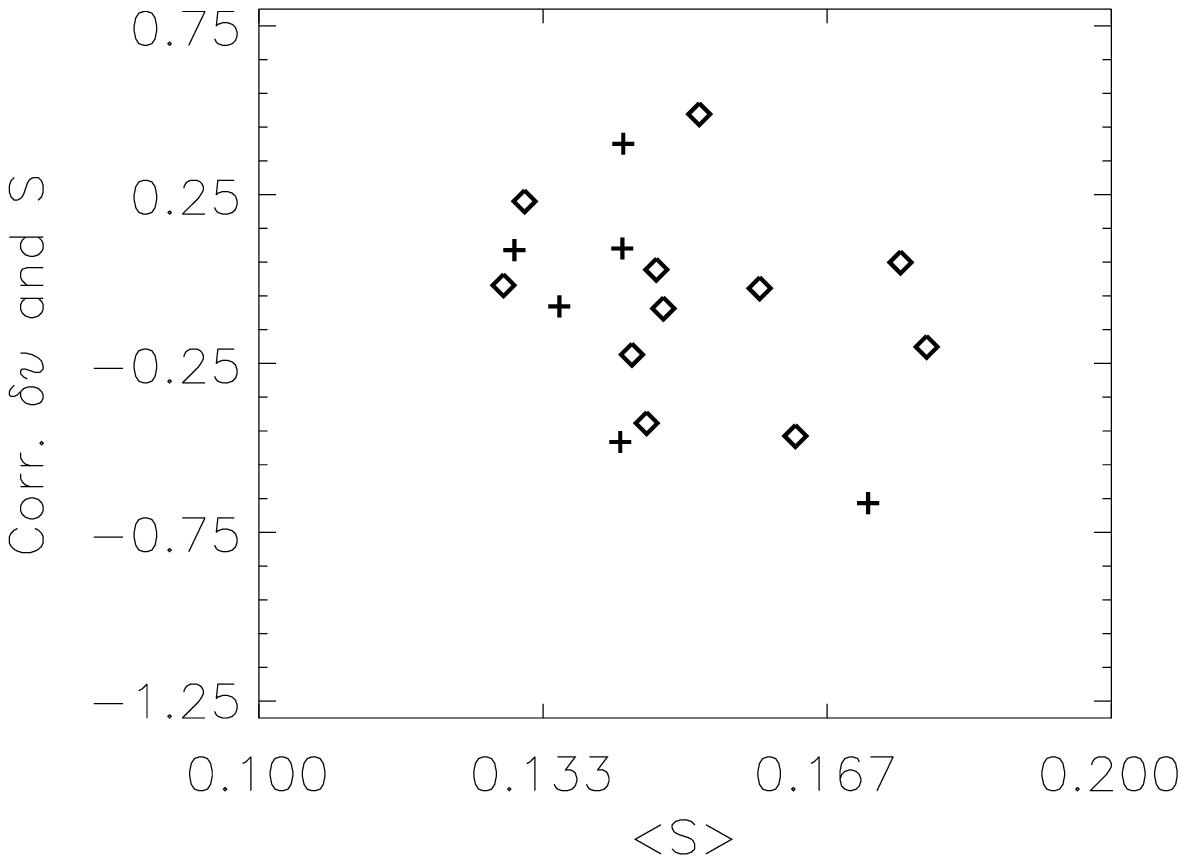}
\caption{Correlation between the eigenmode frequencies shifts and the chromospheric emission as a function of the mean value of the chromospheric emission. Crosses mark G-type dwarfs and diamonds mark F-type dwarfs.}
\end{figure}

\begin{figure}
\includegraphics[width=8cm]{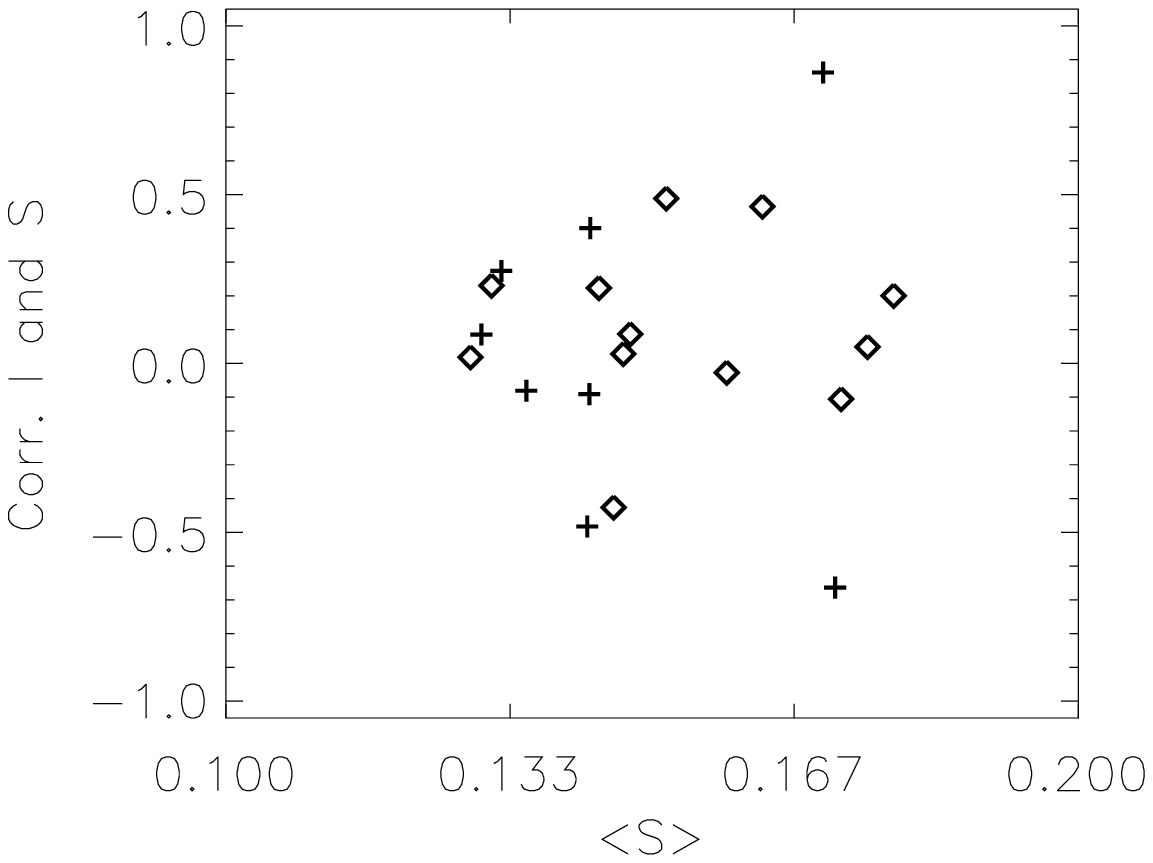}
\caption{Correlation between the relative flux and the chromospheric emission as a function of the mean value of the chromospheric emission. Symbols as in Fig.~6.}
\end{figure}

 We did test all results using $R'_{\rm HK}$ instead of the $S$ index and no significant differences were found.

\section{Discussions}
So far KIC 8006161 is the only star observed by $Kepler$ that is known to show a bona fide cycle comparable to the 11-year solar cycle \citep[see][for a detailed analysis of this star]{2018ApJ...852...46K}. In our analysis it is also the only star to show a correlation above the 95\% significance level between the variability in the chromospheric emission, the eigenmode frequency shifts and the relative flux. 

KIC 10124866 shows a very weak correlation between the chromospheric emission and the relative flux, but the correlation is not above the 95\% significance level. No other star shows correlations between the different activity indicators that are above the 95\% significance level.

KIC~8379927 shows a nice correlation between the variability in the chromospheric emission and the eigenmode frequencies, but the significance is only 61\%. This star is known to be a spectroscopic binary with a $1743.3\pm2.5$ day period \citep{2007Obs...127..313G}. KIC~8379927 was flagged by \citet{2018ApJS..237...17S} as one of the stars that showed evidence of activity-related frequency shifts. Though the significance is very low, we do agree with this interpretation, that between 2010 and early 2012 KIC~8379927 shows a continuous rise and fall in the frequency shifts that is also seen in the chromospheric emission. It is however, not clearly seen in the relative flux, though the measurements in 2011 seem to show larger than average scatter. 
It is still too early to conclude if the variability we see in KIC~8379927 is related to a stellar cycle. Looking at the combined frequency shifts and chromospheric emission, there could be hints of a 2-year periodic variability, with maxima in mid-2009, mid-2011 and early/mid-2013. The time-series are however, too short to make any claims about the significance of such periodic variability.

According to \citet{2017A&A...601A..67C} KIC~8379927 is a Sun-like star with a radius of 1.105 M$_{\odot}$, mass of 1.08 M$_{\odot}$ and age of 1.65 Gyr. According to \citet{2014A&A...572A..34G} the star has a rotation period of $16.99\pm1.35$ days. KIC~8379927 is in other words likely a very Sun-like star and we would therefore expect any periodic variability to be longer than 2 years \citep{2007ApJ...657..486B}. This however, assumes a large mass difference between the two binary components.

Recently, there has been a number of claims of a biennial cycle in the Sun \citep{2010ApJ...718L..19F, 2012MNRAS.420.1405B, 2012A&A...539A.135S, 2013ApJ...765..100S}. What we are seeing in KIC~8379927 could thus be the manifestation of a second dynamo \citep{2010ApJ...718L..19F} comparable to the biennial cycle in the Sun. A number of other G-type dwarfs, like HD 190406, HD 78366 and HD 114710, also show such a secondary biennial cycles \citep{1995ApJ...438..269B, 2016A&A...590A.133O}. Unlike the variability seen in the F-type dwarfs, these secondary biennial cycles seen in the more Sun-like G-type dwarfs could be true polarity reversing solar-like activity cycles \citep{2018MNRAS.479.5266J}. More observations are however, needed before we can make any solid claims. Zeeman-Doppler imaging would be especially useful here, as it can be used to evaluate if we see a polarity flip. 

In general the comparison of the standard deviation of the different activity indicators does not show any correlation (Figs.~3--5). In fact there might be hints of an inverse correlation between the standard deviation of the frequency shifts and the chromospheric emission for stars with $S$ indices with a standard deviation less than 0.0177 (Fig.~3). The correlation does not seem to be related to any effective temperature or metallicity effects. As no good explanation for such an inverse correlation exists, this may simply be due to noise.

In Fig.~4 we see an inverse relation between the standard deviation of the relative flux $\sigma I$ and the chromospheric emission $\sigma S$ for the stars with mean chromospheric emission larger than 0.15. This behaviour could be explained by a scenario where stars slightly more active than the Sun goes from being spot dominated to being faculae dominated and in this transition $\sigma I$ decreases with increasing chromospheric activity. Such a inverse relation would not show up for the inactive stars as they would be expected to have a more constant spot-to-faculae ratio throughout the cycle. This means that any spot darkening is balanced out by faculae brightening. The chromospheric emission is however, not affected by this phenomenon. 

Alternately, the inverse relation between $\sigma I$ and $\sigma S$ could be explained by shifts in the distributions of active regions, where the area covered by spots increases, but the area covered by plage becomes more spatially homogeneous.  These sorts of patterns are seen in more active stars \citep{2018ApJ...855...75R}.

If lack of correlations for the inactive stars is due to a missing degeneracy of the signals form spots and faculae then the problem could be solved by using multi-band photometry, since due to their different temperatures, faculae and spots cannot successfully cancel each other out at all wavelengths.

We do note that we were able in Paper II to see a correlation between the mean value of the excess flux and both the rotation period and age.

We further investigated if the correlation between the different activity indicators depends on the absolute level of the chromospheric emission (Figs~6 \& 7). Again, no correlation was found. It is not unlikely that the missing correlation is due to the sparse sampling of the measurements of the chromospheric emission and the fact that the relative flux measurements have to be interpolated for this comparison. In other words, even though we do not see any strong correlation between the different activity indicators, it is not the same as claiming that they are not there.

A correlation between the $R'_{\rm HK}$ calibration and metallicity has been noted by \citet{2006SPD....37.1201S}, \citet{2008A&A...485..571J} and \citet{2012IAUS..286..335S}. Indicating that metallicity suppress the measured $R'_{\rm HK}$ values. We searched for such a correlation between metallicity and the other measured parameters, but did not find any. Again, this could be due to our limited sample size, but also that our stars span a narrow range in effective temperature.

In general, the main conclusion from our work on {\it sounding stellar cycles with Kepler} is that we need a longer time baseline for the observations. In the most optimal case this means following the 20 targets in our sample and maybe even more stars indefinitely. We did not find such an approach feasible by submitting new observing proposals to the NOT every semester. Instead we have installed an eShell spectrograph system from Shelyak Instruments at the Hertzsprung SONG telescope at Teide Observatory and are at the moment testing if it can be used, with minor modifications, to measure chromospheric activity in our targets. If these test turn out to be successful, we will have the option to monitor the variability of the chromospheric emission in these stars for as long as SONG is operational. We note that the main spectrograph at SONG is optimised for measuring accurate radial velocities with the use of a iodine cell and is therefore optimised for the wave lengths covered by the main iodine lines (4900-6200 {\AA}). This means that the Ca~{\sc ii} $H$ and $K$ lines do not even fall on the detector.

At SONG we would try to follow an observing strategy that comes closer to the observing strategy used in the Mount Wilson project, where the targets were observed weekly, when visible, than the observing strategy we have used, where the targets were only observed three times per semester. We propose to follow such a strategy, even though it would result in fewer observed targets. There are two main reasons for such a prioritisation. Firstly, as we argue that the missing correlation between the different activity indicators in this study is due to the sparse sampling (and resulting interpolation), a higher sampling cadence should result in higher correlations between the different activity indicators and therefore also more reliable measurements. Secondly,  a higher sampling rate would allow a robust mean and standard deviation of the amplitude of the rotational modulation to be calculated. Thirdly, as shown in \citet{2018ApJ...852...46K}, it is possible to use seasonal measured rotation rates to see indications of surface differential rotation caused by active regions at different latitudes. This is however, only possible if sufficient observations are available each semester for determining the rotation period. 

Based on our experience with our NOT program, we recommend that the S/N should be not much less that 100 at the blue part of the spectrum in order to be able to obtain reliable activity measurements.

It is our aim that the new spectrograph for measuring stellar activity with SONG should be fully ready for science observations when NASA's Transiting Exoplanet Survey Satellite (TESS) will start new observations of the $Kepler$ field. This will allow us to test if a denser sampling will lead to a better correlation between the different activity indicators. Apart from this, TESS will provide us with asteroseismic measurements of the fundamental stellar parameters of the stars in which activity cycles have been observed from the Mount Wilson and Lowell observations and will therefore constitute a nice continuation of our {\it sounding stellar cycles with Kepler} program. This is especially true if TESS was extended beyond the nominal mission, which would allow the Kepler field to be revisited each summer.

\section{Conclusions}
We present four years of observations of emission in the Ca~{\sc ii} $H$ and $K$ lines in 20 Sun-like stars. The observations were undertaken simultaneously with observations by the $Kepler$ mission. This has allowed us to analyse the relation between cycle-induced changes on the surface and in the interior in the metal rich Sun-like star KIC 8006161 as seen in the chromospheric emission, in the eigenmode frequencies and in the relative flux \citep{2018ApJ...852...46K}.

The comparison for the different activity indicators for the remaining 19 stars does however, not show the same agreement as seen for KIC 8006161. We suggest that this is mainly due to two effects. Firstly, the sparse sampling of our spectrographic observations make it difficult to compare with the photometric $Kepler$ observations and secondly, due to the failure of $Kepler's$ reaction wheels, we only have 4 years of observations available, which is not sufficient to detect an activity cycle similar to the 11-year solar cycle. We are therefore trying to set up multidecadal observations of the 20 stars with a new dedicated spectrograph at the SONG telescope.

\section*{Acknowledgements}

We would like to thank the referee for thoughtful comments, which significantly improved the paper. The project is been supported by the Villum Foundation. Funding for the Stellar Astrophysics Centre is provided by the Danish National Research Foundation (Grant agreement No.: DNRF106). TSM acknowledges support from a Visiting Fellowship at the Max Planck Institute for Solar System Research. ARGS acknowledges the support from National Aeronautics and Space Administration under Grant NNX17AF27G. The Nordic Optical Telescope is operated by the Nordic Optical Telescope Scientific Association at the Observatorio del Roque de los Muchachos, La Palma, Spain, of the Instituto de Astrof{\' i}sica de Canarias.

\bsp	% typesetting comment
\label{lastpage}

\begin{thebibliography}{99}
\bibitem[Aigrain et al.(2015)]{2015MNRAS.450.3211A} Aigrain, S., Llama, J., Ceillier, T., et al.\ 2015, \mnras, 450, 3211 
\bibitem[Baliunas et al.(1995)]{1995ApJ...438..269B} Baliunas, S.~L., Donahue, R.~A., Soon, W.~H., et al.\ 1995, \apj, 438, 269
\bibitem[Balona \& Abedigamba(2016)]{2016MNRAS.461..497B} Balona, L.~A., \& Abedigamba, O.~P.\ 2016, \mnras, 461, 497 
\bibitem[Balona et al.(2016)]{2016MNRAS.463.1740B} Balona, L.~A., {\v S}vanda, M., \& Karlick{\'y}, M.\ 2016, \mnras, 463, 1740 
\bibitem[Barnes et al.(2016)]{2016AN....337..810B} Barnes, S.~A., Spada, F., \& Weingrill, J.\ 2016, Astronomische Nachrichten, 337, 810 
\bibitem[Basri(2018)]{2018ApJ...865..142B} Basri, G.\ 2018, \apj, 865, 142 
\bibitem[Bazot et al.(2018)]{2018arXiv181008630B} Bazot, M., Nielsen, M.~B., Mary, D., et al.\ 2018, arXiv:1810.08630 
\bibitem[Benomar et al.(2018)]{2018Sci...361.1231B} Benomar, O., Bazot, M., Nielsen, M.~B., et al.\ 2018, Science, 361, 1231 
\bibitem[B{\"o}hm-Vitense(2007)]{2007ApJ...657..486B} B{\"o}hm-Vitense, E.\ 2007, \apj, 657, 486 
\bibitem[Bonanno et al.(2014)]{2014A&A...569A.113B} Bonanno, A., Fr{\"o}hlich, H.-E., Karoff, C., et al.\ 2014, \aap, 569, A113 
\bibitem[Brandenburg et al.(2017)]{2017ApJ...845...79B} Brandenburg, A., Mathur, S., \& Metcalfe, T.~S.\ 2017, \apj, 845, 79 
\bibitem[Broomhall et al.(2012)]{2012MNRAS.420.1405B} Broomhall, A.-M., Chaplin, W.~J., Elsworth, Y., \& Simoniello, R.\ 2012, \mnras, 420, 1405
\bibitem[Bruntt et al.(2012)]{2012MNRAS.423..122B} Bruntt, H., Basu, S., Smalley, B., et al.\ 2012, \mnras, 423, 122 
\bibitem[Christensen-Dalsgaard(2002)]{2002RvMP...74.1073C} Christensen-Dalsgaard, J.\ 2002, Reviews of Modern Physics, 74, 1073 
\bibitem[Ceillier et al.(2017)]{2017A&A...605A.111C} Ceillier, T., Tayar, J., Mathur, S., et al.\ 2017, \aap, 605, A111 
\bibitem[Creevey et al.(2017)]{2017A&A...601A..67C} Creevey, O.~L., Metcalfe, T.~S., Schultheis, M., et al.\ 2017, \aap, 601, A67 
\bibitem[Dziembowski \& Goode(2005)]{2005ApJ...625..548D} Dziembowski, W.~A., \& Goode, P.~R.\ 2005, \apj, 625, 548 
\bibitem[Dziembowski \& Goode(2004)]{2004ApJ...600..464D} Dziembowski, W.~A., \& Goode, P.~R.\ 2004, \apj, 600, 464 
\bibitem[dos Santos et al.(2016)]{2016A&A...592A.156D} dos Santos, L.~A., Mel{\'e}ndez, J., do Nascimento, J.-D., et al.\ 2016, \aap, 592, A156 
\bibitem[Duncan et al.(1991)]{1991ApJS...76..383D} Duncan, D.~K., Vaughan, A.~H., Wilson, O.~C., et al.\ 1991, \apjs, 76, 383 
\bibitem[Eddy(1976)]{1976Sci...192.1189E} Eddy, J.~A.\ 1976, Science, 192, 1189
\bibitem[Ferreira Lopes et al.(2015)]{2015A&A...583A.134F} Ferreira Lopes, C.~E., Le{\~a}o, I.~C., de Freitas, D.~B., et al.\ 2015, \aap, 583, A134 
\bibitem[Fletcher et al.(2010)]{2010ApJ...718L..19F} Fletcher, S.~T., Broomhall, A.-M., Salabert, D., et al.\ 2010, \apjl, 718, L19 
\bibitem[Frandsen \& Lindberg(2000)]{2000mons.proc..163F} Frandsen, S., \& Lindberg, B.\ 2000, The Third MONS Workshop: Science Preparation and Target Selection, 163 
\bibitem[Garc{\'{\i}}a et al.(2010)]{2010Sci...329.1032G} Garc{\'{\i}}a, R.~A., Mathur, S., Salabert, D., et al.\ 2010, Science, 329, 1032 
\bibitem[Garc{\'{\i}}a et al.(2014)]{2014A&A...572A..34G} Garc{\'{\i}}a, R.~A., Ceillier, T., Salabert, D., et al.\ 2014, \aap, 572, A34 
\bibitem[Gizon(2004)]{2004SoPh..224..217G} Gizon, L.\ 2004, \solphys, 224, 217 
\bibitem[Griffin(2007)]{2007Obs...127..313G} Griffin, R.~F.\ 2007, The Observatory, 127, 313 
\bibitem[Grundahl et al.(2017)]{2017ApJ...836..142G} Grundahl, F., Fredslund Andersen, M., Christensen-Dalsgaard, J., et al.\ 2017, \apj, 836, 142 
\bibitem[H{\o}g et al.(2000)]{2000A&A...355L..27H} H{\o}g, E., Fabricius, C., Makarov, V.~V., et al.\ 2000, \aap, 355, L27 
\bibitem[Isaacson \& Fischer(2010)]{2010ApJ...725..875I} Isaacson, H., \& Fischer, D.\ 2010, \apj, 725, 875
\bibitem[Jeffers et al.(2018)]{2018MNRAS.479.5266J} Jeffers, S.~V., Mengel, M., Moutou, C., et al.\ 2018, \mnras, 479, 5266 
\bibitem[Jenkins et al.(2008)]{2008A&A...485..571J} Jenkins, J.~S., Jones, H.~R.~A., Pavlenko, Y., et al.\ 2008, \aap, 485, 571
\bibitem[Judge et al.(2017)]{2017ApJ...848...43J} Judge, P.~G., Egeland, R., Metcalfe, T.~S., Guinan, E., \& Engle, S.\ 2017, \apj, 848, 43 
\bibitem[Karoff et al.(2009)]{2009MNRAS.399..914K} Karoff, C., Metcalfe, T.~S., Chaplin, W.~J., et al.\ 2009, \mnras, 399, 914 
\bibitem[Karoff et al.(2013)]{2013MNRAS.433.3227K} Karoff, C., Metcalfe, T.~S., Chaplin, W.~J., et al.\ 2013, \mnras, 433, 3227 
\bibitem[Karoff et al.(2018)]{2018ApJ...852...46K} Karoff, C., Metcalfe, T.~S., Santos, A.~R.~G., et al.\ 2018, \apj, 852, 46
\bibitem[Kiefer et al.(2017)]{2017A&A...598A..77K} Kiefer, R., Schad, A., Davies, G., \& Roth, M.\ 2017, \aap, 598, A77 
\bibitem[Lovis et al.(2011)]{2011arXiv1107.5325L} Lovis, C., Dumusque, X., Santos, N.~C., et al.\ 2011, arXiv:1107.5325 
\bibitem[Mathur et al.(2013)]{2013A&A...550A..32M} Mathur, S., Garc{\'{\i}}a, R.~A., Morgenthaler, A., et al.\ 2013, \aap, 550, A32 
\bibitem[Mathur et al.(2014)]{2014A&A...562A.124M} Mathur, S., Garc{\'{\i}}a, R.~A., Ballot, J., et al.\ 2014, \aap, 562, A124 
\bibitem[McQuillan et al.(2013)]{2013ApJ...775L..11M} McQuillan, A., Mazeh, T., \& Aigrain, S.\ 2013, \apjl, 775, L11 
\bibitem[McQuillan et al.(2014)]{2014ApJS..211...24M} McQuillan, A., Mazeh, T., \& Aigrain, S.\ 2014, \apjs, 211, 24 
\bibitem[Mehrabi et al.(2017)]{2017ApJ...834..207M} Mehrabi, A., He, H., \& Khosroshahi, H.\ 2017, \apj, 834, 207 
\bibitem[Metcalfe et al.(2016)]{2016ApJ...826L...2M} Metcalfe, T.~S., Egeland, R., \& van Saders, J.\ 2016, \apjl, 826, L2 
\bibitem[Metcalfe \& van Saders(2017)]{2017SoPh..292..126M} Metcalfe, T.~S., \& van Saders, J.\ 2017, \solphys, 292, 126 
\bibitem[Middelkoop(1982)]{1982A&A...107...31M} Middelkoop, F.\ 1982, \aap, 107, 31 
\bibitem[Montet et al.(2017)]{2017ApJ...851..116M} Montet, B.~T., Tovar, G., \& Foreman-Mackey, D.\ 2017, \apj, 851, 116 
\bibitem[Nielsen et al.(2013)]{2013A&A...557L..10N} Nielsen, M.~B., Gizon, L., Schunker, H., \& Karoff, C.\ 2013, \aap, 557, L10 
\bibitem[Noyes et al.(1984)]{1984ApJ...279..763N} Noyes, R.~W., Hartmann, L.~W., Baliunas, S.~L., Duncan, D.~K., \& Vaughan, A.~H.\ 1984, \apj, 279, 763
\bibitem[Ol{\'a}h et al.(2016)]{2016A&A...590A.133O} Ol{\'a}h, K., K{\H o}v{\'a}ri, Z., Petrovay, K., et al.\ 2016, \aap, 590, A133 
\bibitem[Radick et al.(2018)]{2018ApJ...855...75R} Radick, R.~R., Lockwood, G.~W., Henry, G.~W., Hall, J.~C., \& Pevtsov, A.~A.\ 2018, \apj, 855, 75 
\bibitem[R{\'e}gulo et al.(2016)]{2016A&A...589A.103R} R{\'e}gulo, C., Garc{\'{\i}}a, R.~A., \& Ballot, J.\ 2016, \aap, 589, A103 
\bibitem[Reinhold et al.(2013)]{2013A&A...560A...4R} Reinhold, T., Reiners, A., \& Basri, G.\ 2013, \aap, 560, A4 
\bibitem[Reinhold \& Gizon(2015)]{2015A&A...583A..65R} Reinhold, T., \& Gizon, L.\ 2015, \aap, 583, A65 
\bibitem[Reinhold et al.(2017)]{2017A&A...603A..52R} Reinhold, T., Cameron, R.~H., \& Gizon, L.\ 2017, \aap, 603, A52
\bibitem[Rutten(1984)]{1984A&A...130..353R} Rutten, R.~G.~M.\ 1984, \aap, 130, 353 
\bibitem[Saar(2006)]{2006SPD....37.1201S} Saar, S.~H.\ 2006, Bulletin of the American Astronomical Society, 38, 12.01 
\bibitem[Saar \& Testa(2012)]{2012IAUS..286..335S} Saar, S.~H., \& Testa, P.\ 2012, Comparative Magnetic Minima: Characterizing Quiet Times in the Sun and Stars, 286, 335 
\bibitem[Salabert et al.(2016)]{2016A&A...596A..31S} Salabert, D., Garc{\'{\i}}a, R.~A., Beck, P.~G., et al.\ 2016, \aap, 596, A31 
\bibitem[Salabert et al.(2018)]{2018A&A...611A..84S} Salabert, D., R{\'e}gulo, C., P{\'e}rez Hern{\'a}ndez, F., \& Garc{\'{\i}}a, R.~A.\ 2018, \aap, 611, A84 
\bibitem[Santos et al.(2018)]{2018ApJS..237...17S} Santos, A.~R.~G., Campante, T.~L., Chaplin, W.~J., et al.\ 2018, \apjs, 237, 17 
\bibitem[Schrijver \& Zwaan(2008)]{2008ssma.book.....S} Schrijver, C.~J., \& Zwaan, C.\ 2008, Solar and Stellar Magnetic Activity, by C.~J.~Schrijver , C.~Zwaan, Cambridge, UK: Cambridge University Press
\bibitem[Shapiro et al.(2017)]{2017NatAs...1..612S} Shapiro, A.~I., Solanki, S.~K., Krivova, N.~A., et al.\ 2017, Nature Astronomy, 1, 612 
\bibitem[Simoniello et al.(2012)]{2012A&A...539A.135S} Simoniello, R., Finsterle, W., Salabert, D., et al.\ 2012, \aap, 539, A135 
\bibitem[Simoniello et al.(2013)]{2013ApJ...765..100S} Simoniello, R., Jain, K., Tripathy, S.~C., et al.\ 2013, \apj, 765, 100 
\bibitem[Solanki et al.(2013)]{2013ARA&A..51..311S} Solanki, S.~K., Krivova, N.~A., \& Haigh, J.~D.\ 2013, \araa, 51, 311 
\bibitem[van Saders et al.(2016)]{2016Natur.529..181V} van Saders, J.~L., Ceillier, T., Metcalfe, T.~S., et al.\ 2016, \nat, 529, 181 
\bibitem[Wright(2005)]{2005PASP..117..657W} Wright, J.~T.\ 2005, \pasp, 117, 657 
\end{thebibliography}
\end{document}